\documentclass[twocolumn]{aastex631}
\usepackage{multirow}
\usepackage{amsmath}
\shorttitle{Radio Phoenix in the Abell 85}
\shortauthors{Raja et al.}
\begin{document}

\title{A Multi-Frequency View of the Radio Phoenix in the Abell 85 Cluster}

\correspondingauthor{Ramij Raja}
\email{ramij.amu48@gmail.com}

\author[0000-0001-8721-9897]{Ramij Raja}
\affiliation{Centre for Radio Astronomy Techniques and Technologies, Department of Physics and Electronics,\\ Rhodes University, Makhanda 6140, South Africa}

\author[0000-0002-1372-6017]{Majidul Rahaman}
\affiliation{Institute of Astronomy, National Tsing Hua University, Hsinchu 300044, Taiwan, R.O.C.}

\author[0000-0002-5333-1095]{Abhirup Datta}
\affiliation{Department of Astronomy, Astrophysics and Space Engineering, Indian Institute of Technology Indore, Simrol, 453552, India}

\author[0000-0003-1680-7936]{Oleg M. Smirnov}
\affiliation{Centre for Radio Astronomy Techniques and Technologies, Department of Physics and Electronics,\\ Rhodes University, Makhanda 6140, South Africa}
\affiliation{South African Radio Astronomy Observatory, 2 Fir Street, Black River Park, Observatory, Cape Town 7925, South Africa}



\begin{abstract}

Radio phoenices are complex and filamentary diffuse radio sources found in both merging and relaxed clusters. The formation of these sources was proposed to be due to adiabatic compression of old Active Galactic Nucleus (AGN) plasma in shock waves. Most of the previous spectral studies of these sources were limited to integrated spectral indices, which were found to be very steep as well as showing a curved spectrum. Here, we have performed a multi-frequency investigation of the radio phoenix in the Abell 85 cluster. Owing to the sensitive high-resolution observations, we found some of the finer filamentary structures that were previously undetected. We produced resolved spectral index maps of the radio phoenix between 323, 700, and 1280 MHz. The orientation of the filaments, as well as the gradient across the spectral index maps, suggest the possible direction of the shock motion from northeast to southwest. 
The integrated spectrum of the radio phoenix was found to be very steep and curved toward high-frequencies.
Furthermore, the spectral index of the filaments was found to be less steep compared to the non-filamentary regions, implying greater energy injection in the filaments. The observed features in the radio phoenix in the Abell 85 cluster seem to be in support of the adiabatic shock compression mechanism.

\end{abstract}



\section{Introduction} \label{sec:intro}

Galaxy clusters are the largest gravitationally bound objects in the present Universe and also act as a fair sample to study the structure formation processes in the Universe. We have come a long way since the first discovery of diffuse radio synchrotron emission from the Coma cluster through radio observations by \citet{Large1959Natur.183.1663L}. Especially the last few decades have seen tremendous advances in finding new diffuse radio objects in the galaxy clusters \citep[see ][for observational review]{Feretti2012A&ARv..20...54F,vanWeeren2019SSRv..215...16V,Paul2023JApA...44...38P} as well as the development of a theoretical framework for explaining their formation processes \citep[see ][for a theoretical review]{Brunetti2014IJMPD..2330007B,Paul2023JApA...44...38P}. 

Galaxy cluster diffuse radio objects are historically classified as radio halos, relics, and minihalos. However, recent advances in sensitive low-frequency radio observations have uncovered a lot of complex and intermediate-class objects \citep{vanWeeren2011A&A...527A.114V,deGasperin2017SciA....3E1634D,vanWeeren2017NatAs...1E...5V,Raja2020MNRAS.493L..28R}. \citet{vanWeeren2019SSRv..215...16V} proposed a new classification scheme where the whole zoo of cluster diffuse radio objects are divided into three broad classes: radio halos, radio relics, and revived AGN fossil plasma sources. Here, radio halos include both classical giant radio halos \citep{,Shimwell2014MNRAS.440.2901S,Pearce2017ApJ...845...81P,Rajpurohit2023A&A...669A...1R} and radio minihalos \citep{Giacintucci2014ApJ...781....9G,Gendron-Marsolais2017MNRAS.469.3872G,Raja2020ApJ...889..128R}. They typically have regular morphology and are located in the cluster center. The typical size of halos and minihalos are $\sim 1-2$ Mpc and $\sim 0.1-0.5$ Mpc, respectively. 
Radio relics, on the other hand, are of $\sim 1-2$ Mpc scale elongated radio objects located in the cluster peripheral regions. They appear filamentary when viewed at high resolution \citep{vanWeeren2016ApJ...818..204V,DiGennaro2018ApJ...865...24D,Rajpurohit2022ApJ...927...80R}. The smaller in size and located comparatively nearer to the cluster center, radio phoenices \citep{Kempner2004rcfg.proc..335K} were previously classified along with the radio relics \citep[e.g.,][]{Slee2001AJ....122.1172S}, but are now classified along with other revived fossil plasma sources \citep{deGasperin2017SciA....3E1634D}. 

Radio phoenices are diffuse radio objects that come in a variety of shapes. Their morphology varies from roundish \citep[e.g., Abell 1664, ][]{Giovannini1999NewA....4..141G,Kale2012ApJ...744...46K} to elongated and filamentary \citep[e.g., Abell 2593, ][]{Mandal2020A&A...634A...4M}. They are found in both merging \citep[e.g., Abell 2256, ][]{vanWeeren2009A&A...508.1269V} and relaxed \citep[e.g., Abell 4038, ][]{Kale2018MNRAS.480.5352K} clusters. They are located in between the cluster center and peripheral region. The characteristic features of these sources are their linear extents of hundreds of kpc \citep{deGasperin2015MNRAS.448.2197D} with AGN origin of the plasma \citep{Kale2018MNRAS.480.5352K}, and steep spectra due to synchrotron loss \citep{vanWeeren2011A&A...527A.114V}. 

Although a few of these diffuse radio sources have been discovered so far, a detailed analysis of their origin is still lacking. 
Some of the notable works on the spectral analysis of radio phoenices are \citet{vanWeeren2011A&A...527A.114V,Kale2012ApJ...744...46K,deGasperin2015MNRAS.448.2197D,Kale2018MNRAS.480.5352K,Mandal2020A&A...634A...4M}. 
To explain the diffuse synchrotron emission from fossil plasma in galaxy clusters, \citet{Enblin2001A&A...366...26E} proposed a scenario where a merger shock passes through the fossil plasma from past AGN activity, compresses the plasma, resulting in re-energization of the electrons which emit the observed synchrotron emission. This is the current favored scenario to explain the radio phoenices \citep{Kale2012ApJ...744...46K,Mandal2020A&A...634A...4M}. 
Recently, \citet{Rahaman2022MNRAS.515.2245R} studied the Abell 85 cluster in both radio and X-ray and suggested that the radio phoenix was either being formed by the shock present at the eastern edge of the phoenix or the sloshing arm of the intracluster medium (ICM).

In this paper, we have investigated the radio phoenix in Abell 85 using multiple radio frequencies in an attempt to compare observational results with the proposed adiabatic shock compression formation mechanism mentioned above. In the following, we briefly present the previous work on this cluster in Sect. \ref{sec:a85}. The details of the observation and data reduction procedure are discussed in Sect. \ref{sec:obs}. The observational features are discussed in Sect. \ref{sec:radio_phoenix}. The spectral study of the phoenix is presented in Sect. \ref{sec:spectral_study}. Shock properties associated with the phoenix are derived in Sect. \ref{sec:shock_prop}.
A comparison of the radio phoenix with the model is discussed in Sect. \ref{sec:discus}, and finally the conclusions of this study are presented in Sect. \ref{sec:conclude}.

%%%%%%%%%%%%%%%%%%%%%%%%%%%%%%%%%%%%%%%%%%%%%
\section{The Abell 85 cluster} \label{sec:a85}
Abell 85 \citep[$z=0.0556$;][]{Rines2016ApJ...819...63R} is a complex system with a lot of ongoing interaction activity yet it harbors a cool core. Among the two subcluster mergers, one is from the southwest direction and the other is from the south of the cluster center \citep[][]{Durret2005A&A...432..809D,Kempner2002ApJ...579..236K}. \citet{Yu2016ApJ...831..156Y} reported multiple galaxy groups passing through the cluster center, interacting with the ICM in the process. All these features make it an interesting candidate to study dynamical processes in galaxy clusters. 

A lot of studies with radio observations at multiple frequencies have been done on this cluster to understand the diffuse radio emission in question \citep[e.g., ][]{Slee1984PASAu...5..516S,Bagchi1998MNRAS.296L..23B,Giovannini1999NewA....4..141G, Giovannini2000NewA....5..335G,Slee2001AJ....122.1172S,Duchesne2021PASA...38...10D}. 
\citet{Enblin2001A&A...366...26E} presented this radio source as a relic, as an example for their re-acceleration mechanism through fossil plasma compression due to merger shock. \citet{Slee2001AJ....122.1172S} proposed two scenarios: (1) if the relic is associated with the Brightest Cluster Galaxy (BCG) then only recompression proposed by \citet{Enblin2001A&A...366...26E} can explain its origin, (2) the relic is associated with a nearby radio galaxy which provides the seed electrons for the traditional spectral aging models (e.g., Komissarov-Gubanov-Kardashev-Pacholczyk \citealt{Komissarov1994A&A...285...27K}, Murgia-Jaffe-Perola \citealt{Murgia2001A&A...380..102M}) to work. It was \citet{Kempner2004rcfg.proc..335K} who first classified this extended emission in Abell 85 as a radio phoenix, supporting compression mechanism developed by \citet{Enblin2001A&A...366...26E} and \citet{Enblin2002MNRAS.331.1011E}. 

A detailed X-ray study of Abell 85 performed by \citet{Ichinohe2015MNRAS.448.2971I} reported the presence of a possible shock in front of the merging southwest subcluster with a Mach number of 1.4. The ambiguity arises from the position being the interface between the sloshing gas and the merging subcluster. A later work by \citet{Rahaman2022MNRAS.515.2245R} re-confirmed the presence of shock as well as sloshing motion in the ICM. They discussed both the adiabatic shock compression and sloshing motion as a possible player in the phoenix formation. On another note, the latest MeerKAT Galaxy Cluster Legacy Survey \citep{Knowles2022A&A...657A..56K} has uncovered a minihalo at the center of this cluster and \citet{Raja2023MNRAS.526L..70R} discovered a radio bridge connecting the minihalo and the phoenix, which were previously undetected.

Here, we have explored the possibility of the radio phoenix being revived after passing a merging shock other than the one associated with the southwest subcluster. 
Throughout this paper, we adopt a $\Lambda$CDM cosmology with $H_0 = 70$ km s$^{-1}$ Mpc$^{-1}$, $\Omega_{\mathrm{m}} = 0.3$ and $\Omega_\Lambda = 0.7$. At the cluster redshift $z = 0.0556$, $1\arcsec$ corresponds to a physical scale of 1.08 kpc.

% Obs Table
\begin{deluxetable*}{lccccccc}[!t]
\tablecaption{Radio observations of Abell 85 cluster \label{tab:obs}}
\tablecolumns{8}
\tablewidth{0pt}
\tablehead{
\colhead{Array} & \colhead{Project} & \colhead{Frequency} & \colhead{Bandwidth} & \colhead{Channels} & \colhead{Obs. date} & \colhead{Obs. time} & \colhead{PI}\\
\colhead{} & \colhead{} & \colhead{(GHz)} & \colhead{(MHz)} & \colhead{} & \colhead{} & \colhead{(hrs)} & \colhead{}}
\startdata
GMRT & 14PUL01 & 0.148 & 16 & 128 & 1 Sept 2008 & $\sim5$ & Ue Li Pen \\
%\noalign{\smallskip}
GMRT & 14CPA01 & 0.148 & 16 & 128 & 3 Sept 2008 & $\sim5.4$ & Christoph Pfrommer \\
%\noalign{\smallskip}
GMRT & 30\_085 & 0.323 & 33 & 512 & 29 Jul 2016 & $\sim10$ & Stephen Hamer \\
%\noalign{\smallskip}
uGMRT & 39\_104 & 0.7 & 200 & 8192 & 20 Nov 2020 & $\sim11$ & Ramij Raja \\
%\noalign{\smallskip}
MeerKAT & SSV-20180624-FC-01 & 1.28 & 800 & 4096 & 26 Sept 2018 & $\sim8$ & Sharmila \\
\enddata
\end{deluxetable*}

%%%%%%%%%%%%%%%%%%%%%%%%%%%%%%%%%%%%%%%%%%%%%%%%%%%%%%%%%%%%%%%%%
\section{Observations and Data Analysis} \label{sec:obs}

In this study, we have used 148 \& 323 MHz GMRT and 1.28 GHz MeerKAT archival data along with our recent 700 MHz GMRT observation. 
The observations at 148 and 323 were made with the old GMRT, whereas the 700 MHz observation was performed with the upgraded GMRT. 
Details of these observations are presented in Table \ref{tab:obs}. A brief description of the data reduction process for both the GMRT and MeerKAT data is discussed below.

\begin{figure*}[!t]
    \centering
\vspace{-0.4in}
    \includegraphics[width=2\columnwidth]{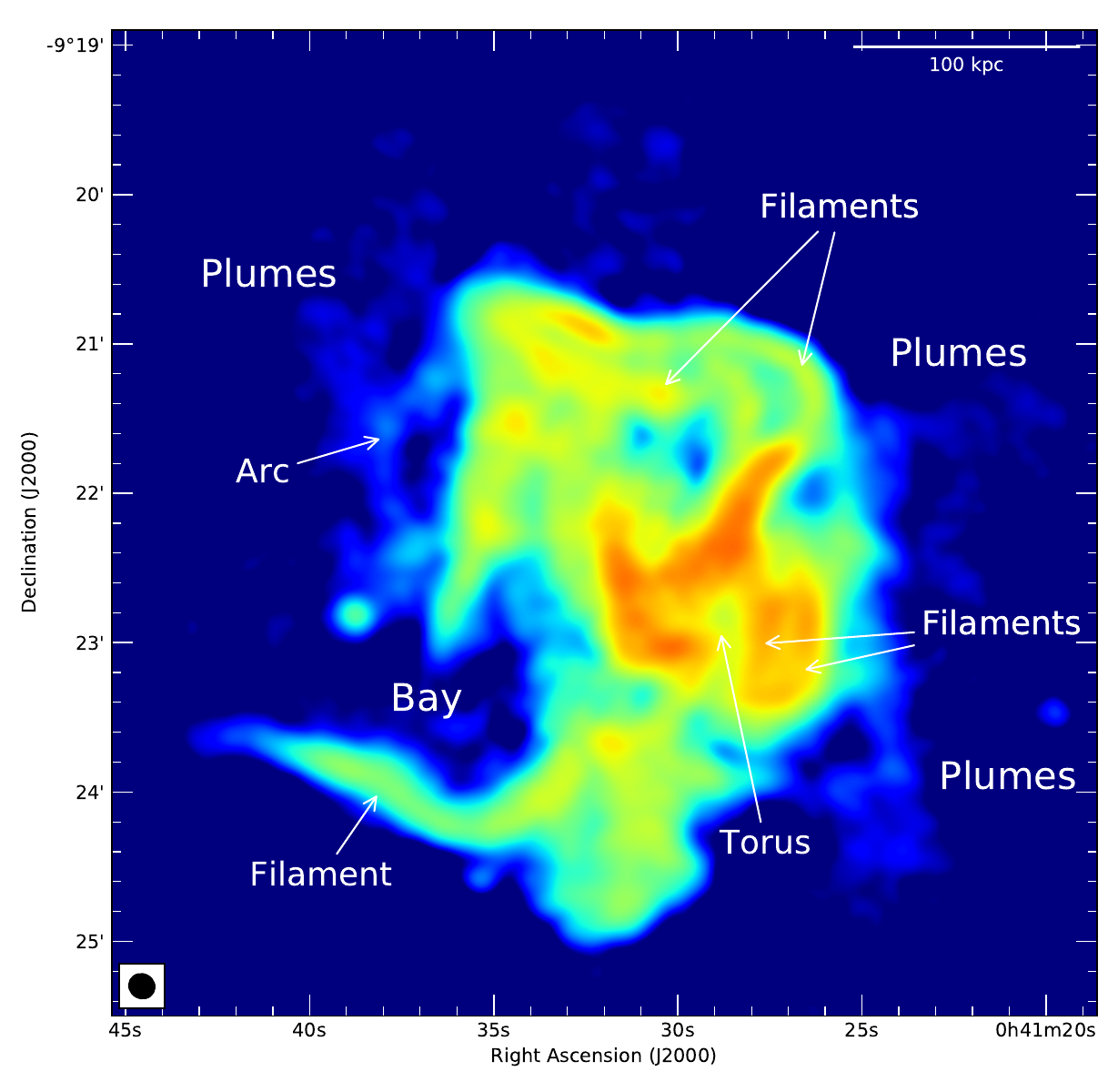}
    
    \caption{High-resolution GMRT radio map of the Abell 85 radio phoenix at 323 MHz. The image properties are given in Table \ref{tab:image_prop}, see label ``IM3''.}
    \label{fig:325MHz_radio_map}
\end{figure*}

\begin{figure*}
    \centering

    \includegraphics[width=2.1\columnwidth]{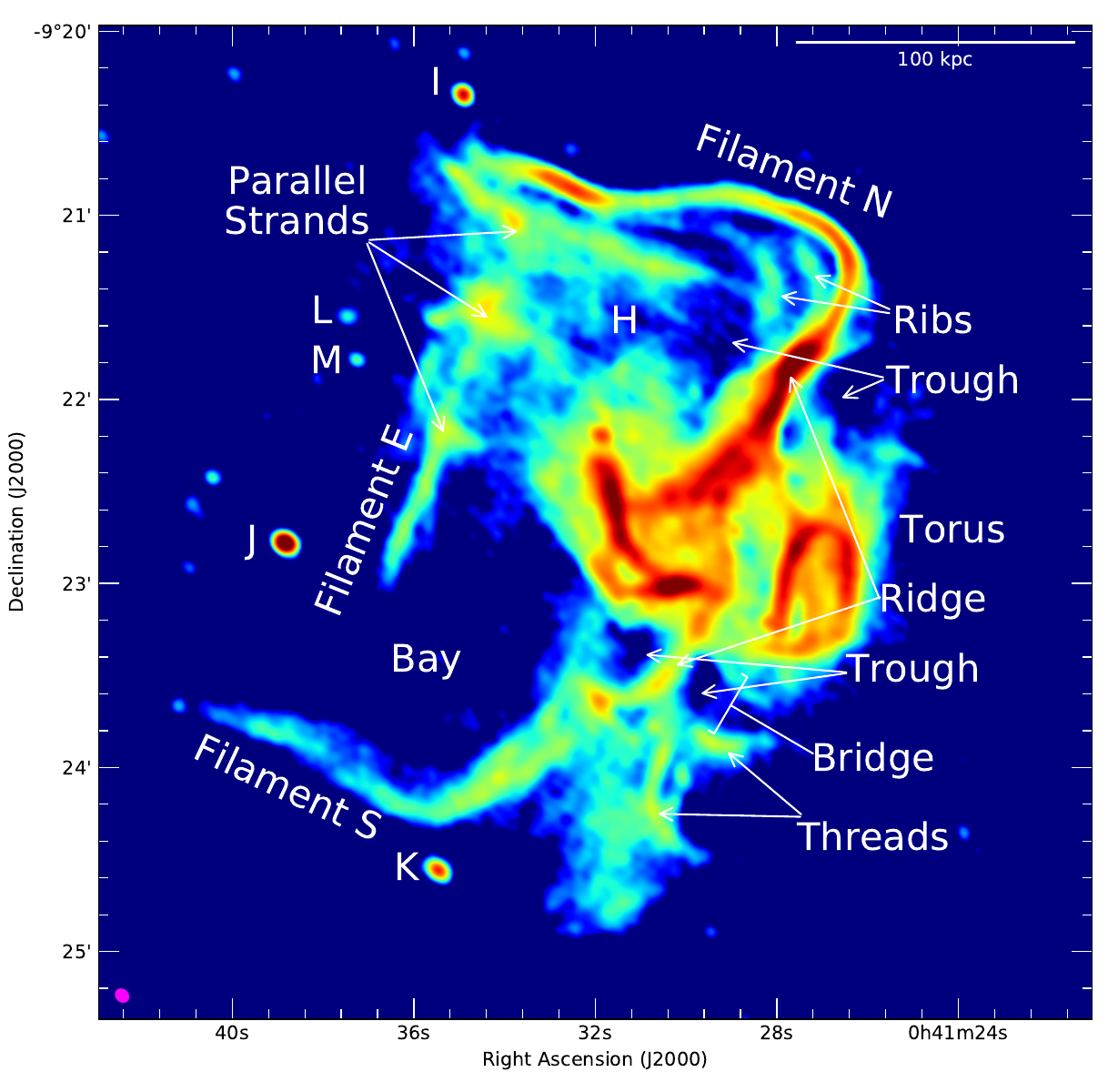}
    
    \caption{High-resolution uGMRT radio map of the Abell 85 radio phoenix at 700 MHz. The image properties are given in Table \ref{tab:image_prop}, see label ``IM7''.}
    \label{fig:700MHz_radio_map}
\end{figure*}

\begin{figure*}[!ht]
    \centering
    \vspace{-0.4in}
    \includegraphics[width=2\columnwidth]{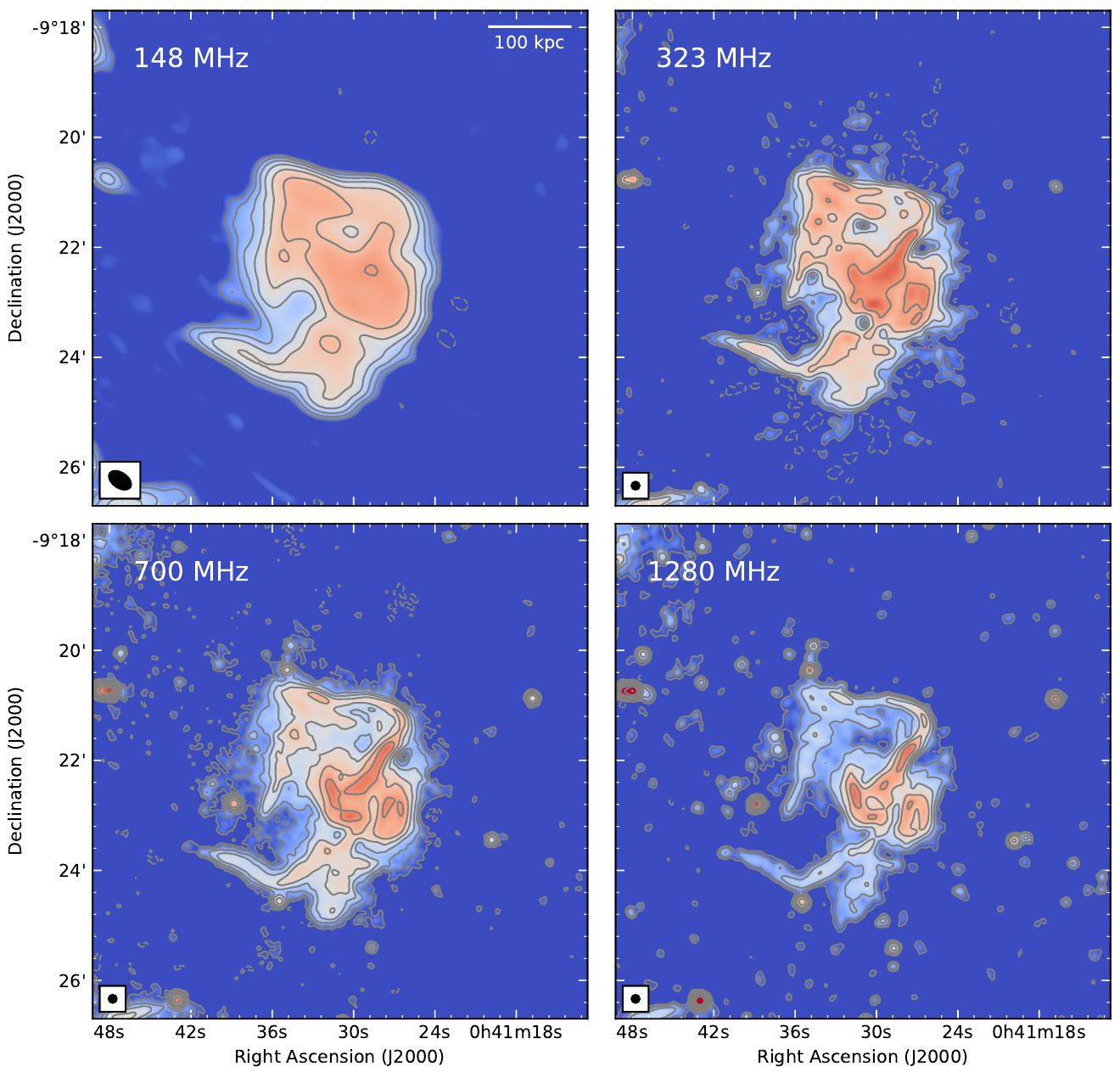}
    \caption{Radio images overlaid with contours of the Abell 85 cluster at 148, 323, 700, and 1280 MHz. The restoring beam of the 148 MHz image is  $\mathrm{Beam_{148}} = 27.3\arcsec \times 16.5\arcsec$, whereas the rest are $9\arcsec$. The radio contours are drawn at levels $[-1,1,2,4,8,...]\times 3\sigma_{\mathrm{rms}}$. The detailed image properties are given in Table \ref{tab:image_prop}. Here, the \textit{top left}, \textit{top right}, \textit{bottom left}, and \textit{bottom right} panels correspond to labels IM1, IM4, IM8, and IM10, respectively.}
    \label{fig:radio_maps}
\end{figure*}

\begin{figure*}
    \centering
    \includegraphics[width=2.1\columnwidth]{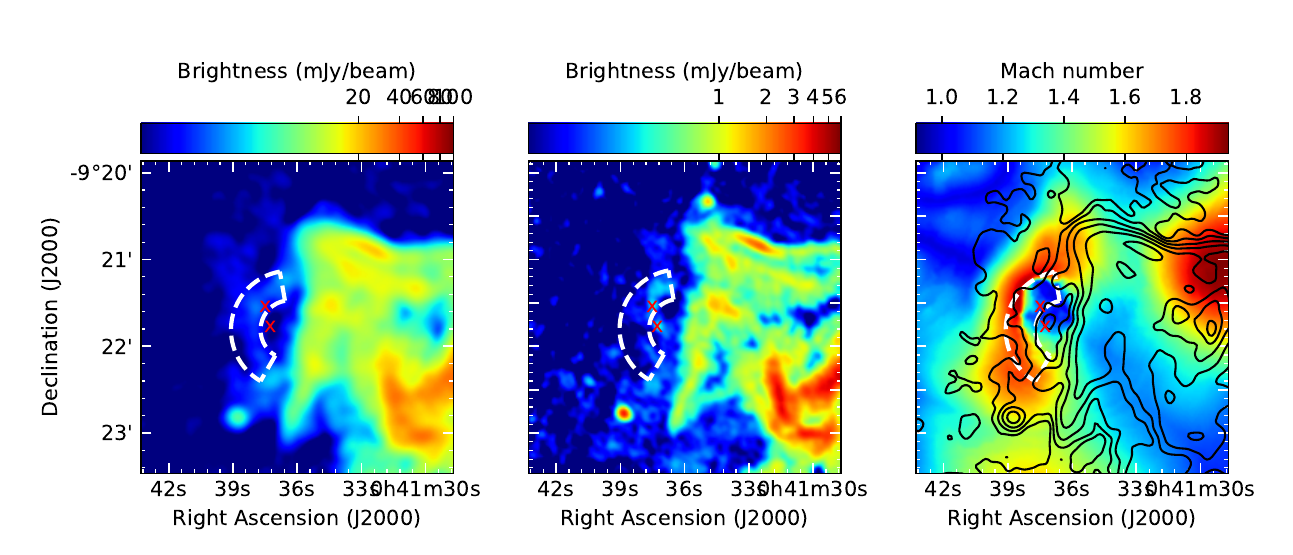}
    \caption{A comparison of 323 MHz image (\textit{left}), 608 MHz image (\textit{middle}), and a Mach number map (\textit{right}) of Abell 85. The white dashed lines outline the `Arc' and the red `X' marks the positions of two compact radio sources. The radio contours overlaid (\textit{right}) are of the 323 MHz image (\textit{left}) and are drawn at levels $[-1,1,2,4,8,...]\times 3\sigma_{\mathrm{rms}}$. Here, the \textit{left} and \textit{middle} panel images correspond to labels IM3 and IM6, respectively, in Table \ref{tab:image_prop}.}
    \label{fig:Arc}
\end{figure*}

\begin{figure*}
    \centering
    \begin{tabular}{ccc}
    \includegraphics[width=0.66\columnwidth]{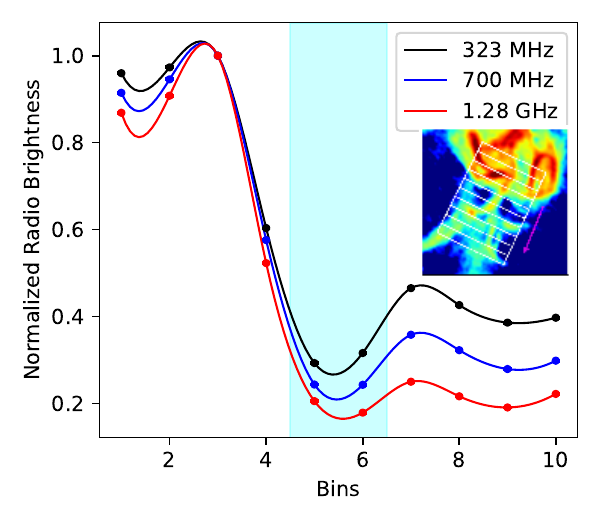} &
    \includegraphics[width=0.66\columnwidth]{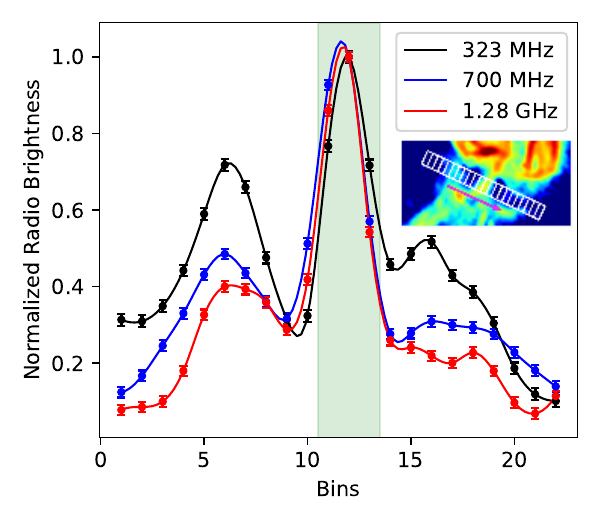} &
    \includegraphics[width=0.66\columnwidth]{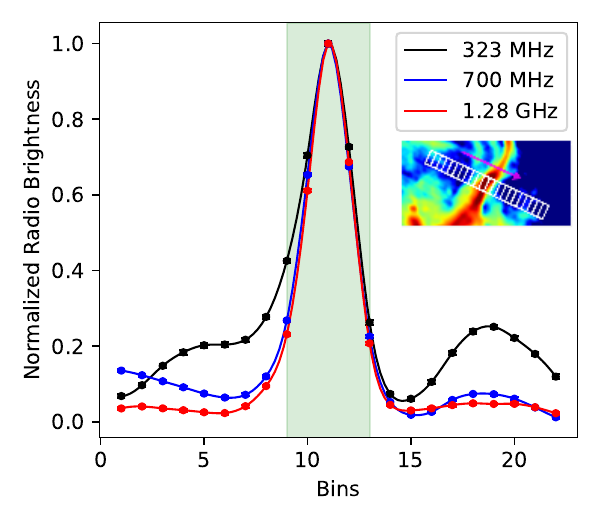} \\
    \end{tabular}
    \caption{Profiles of the `Bridge' region along two different directions. The profiles are normalized with respective peak brightness values. The magenta arrow indicates the plot direction. The shaded regions indicate the approximate width of the `Bridge' filament. The bin widths of the profiles are $10\arcsec$ (\textit{left panel}) and $5\arcsec$ (\textit{middle} \& \textit{right panel}). These profiles are extracted from images IM4, IM8 and IM10.}
    \label{fig:Bridge_profile}
\end{figure*}

%%% Table
\begin{deluxetable*}{lcccccc}%[!t]
\tablecaption{Image properties \label{tab:image_prop}}
\tablecolumns{7}
\tablewidth{0pt}
\tablehead{
\colhead{Array} & \colhead{Frequency} & \colhead{Name} & \colhead{Restoring beam} & \colhead{Briggs} & \colhead{\textit{uv}-range} & \colhead{rms}\\ [-1ex]
\colhead{} & \colhead{(MHz)} & \colhead{} & \colhead{} & \colhead{robust} & \colhead{(k$\lambda$)} & \colhead{($\mu$Jy beam$^{-1}$)}}
\startdata
GMRT & 148 & IM1 & $27.3\arcsec \times 16.5\arcsec$ & 0  & $0.03 - 13.5$ & $2000$ \\
 & & IM2 & $27\arcsec \times 17\arcsec$ & 0  & $0.1 - 13.5$ & $2200$ \\
 \hline
%\noalign{\smallskip}
GMRT & 323 & IM3 & $10.4\arcsec \times 9.6\arcsec$ & 0 & $0.05 - 29$ & $158$ \\
 & & IM4 & $9\arcsec \times 9\arcsec$ & 0 & $0.1 - 29$ & $150$ \\
%  & & IM5 & $7.1\arcsec \times 6.5\arcsec$ & -1 & $2 - 29$ & $100??$ \\
   & & IM5 & $27\arcsec \times 17\arcsec$ & 0 & $0.1 - 13.5$ & $230$ \\
\hline
%\noalign{\smallskip}
GMRT & 608 & IM6 & $6\arcsec \times 5.2\arcsec$ & 0  & $0.1 - 54$ & $44$ \\
\hline
%\noalign{\smallskip}
uGMRT & 700 & IM7 & $4.5\arcsec \times 3.7\arcsec$ & 0 & $0.1 - 74$ & $16$ \\
 & & IM8 & $9\arcsec \times 9\arcsec$ & 0 & $0.1 - 29$ & $20$ \\
 & & IM9 & $27\arcsec \times 17\arcsec$ & 0 & $0.1 - 13.5$ & $51$ \\
\hline
MeerKAT & 1280 & IM10 & $9\arcsec \times 9\arcsec$ & 0 & $0.1 - 29$ & $10$ \\
 & & IM11 & $27\arcsec \times 17\arcsec$ & 0 & $0.1 - 13.5$ & $33$ \\
\enddata
\tablecomments{Images IM1, IM3 and IM6 are produced in \texttt{SPAM}\footnote{\texttt{SPAM} uses \texttt{AIPS} for data processing. The robust parameter is in the \texttt{-5} to \texttt{5} scale, where \texttt{-5} is pure uniform and \texttt{5} is natural weighting.} and the rest are made using \texttt{WSCLEAN}\footnote{In \texttt{WSCLEAN}, the typical robust values are between \texttt{-2} and \texttt{2}, they correspond to the uniform and natural weighting, respectively.}.}
\end{deluxetable*}

%%%%%%
\subsection{GMRT data reduction} \label{subsec:GMRT_data}
For 148 MHz GMRT data reduction, we used the \texttt{SPAM} \citep[Source Peeling and Atmospheric Modeling; ][]{Intema2009A&A...501.1185I,Intema2017A&A...598A..78I} package. This is a python-based pipeline employing \texttt{AIPS} for calibration and imaging. It performs RFI (Radio Frequency Interference) flagging, bandpass, and gain calibration, then proceeds to a few rounds of self-calibration followed by direction-dependent calibration \citep[for details see][]{Intema2017A&A...598A..78I}.

The initial flagging and calibration of the two data sets were done separately and the data was combined during self-calibration and direction-dependent calibration loops in the \texttt{SPAM} pipeline. 
The calibrated data was used for spectral map-making in \texttt{CASA} \citep[Common Astronomy Software Applications;][]{McMullin2007ASPC..376..127M}. 
The 323 MHz data was also processed similarly as discussed above.
The 700 MHz uGMRT wide-band data reduction was done with the updated \texttt{SPAM} package which splits the wide-band data into multiple narrow-band chunks and processes them in a similar fashion as described above.

\subsection{MeerKAT data reduction} \label{subsec:meerkat_data}
For data reduction purposes of the MeerKAT L-band data, we used the \texttt{CARACal}\footnote{https://caracal.readthedocs.io/en/latest/} pipeline \citep{Jozsa2020}. \texttt{CARACal} is python-based pipeline that allows users to conveniently use different radio astronomy packages (\texttt{CASA}, \texttt{TRICOLOUR}\footnote{https://github.com/ratt-ru/tricolour}, \texttt{WSCLEAN} \citealt{offringa-wsclean-2014,Offringa2017MNRAS.471..301O}, \texttt{CUBICAL}\footnote{https://cubical.readthedocs.io/en/latest/}, etc.) for data editing, calibration, and imaging in a scripted format. First, we edited the calibrator data with \texttt{TRICOLOUR} and proceeded with standard \texttt{CASA} calibration steps. To reduce the data volume we averaged over five channels. After a second round of data flagging we proceeded to the self-calibration loops where \texttt{CUBICAL} is used for calibration and \texttt{WSCLEAN} for imaging. After a few rounds of phase-only selfcal, we got the calibrated target data, which was used for later imaging.

%%%%%%%%%%%%%%%%%%%%%%%%%%%%%%%%%%%%%%%%%%%%%%%%%%%%%
\section{The radio phoenix} \label{sec:radio_phoenix}
For a thorough investigation of the radio phoenix, we have produced multi-frequency images corresponding to different \textit{uv}-ranges. 
Among these, the full extent of the radio phoenix is detected only at 323 MHz (IM3; Fig. \ref{fig:325MHz_radio_map}), and the most detailed view of the phoenix was obtained at 700 MHz (IM7; Fig. \ref{fig:700MHz_radio_map}). 
Properties of all images used in this work are presented in Table \ref{tab:image_prop} for ease of reference. All the images used in this work are produced setting Briggs weighting \texttt{robust = 0} to get a good balance of large-scale sensitivity and angular resolution.

%%%%%%%%%%%
\subsection{323 MHz image} \label{subsec:323image}
In Fig. \ref{fig:325MHz_radio_map}, the brightness distribution of the radio phoenix is shown in great detail. The overall structure of the diffuse object is similar to the radio map presented by \citet{Giovannini2000NewA....5..335G}. However, a lot of detailed structures are visible here that are not present in their image. The central part has higher brightness compared to the outer part of this object. Furthermore, a lot of filamentary structures can be seen in this sensitive high-resolution image than that of the previous authors \citep[e.g.,][]{Giovannini2000NewA....5..335G,Slee2001AJ....122.1172S,Duchesne2021PASA...38...10D}{}{}. Some of the distinguishing features of this image are that the northern filament consists of two parallel strands, even bending in a similar way. The central torus is made up of multiple sub-filaments. Weird plume-like structures are present around the main body of the phoenix, these might be related to gas diffusion. Because of low surface brightness, these `plumes' are only observed at the low frequency and are absent at higher frequencies (Fig. \ref{fig:radio_maps}). A less bright filament in the shape of an `Arc' is observed at the north-eastern part of the phoenix. Finally, the `Bay' region is almost devoid of radio emission. A higher-resolution image at 700 MHz with a lot more features is presented below.

%%%%%%%%%%%%%%%
\subsection{700 MHz image} \label{subsec:700image} 
In Fig. \ref{fig:700MHz_radio_map}, although the overall characteristic features of the phoenix remain the same, in this image we uncover that most of the components are made up of thinner filaments that are obscured in the previous image because of low resolution. 
We see that the central torus consists of two components. One is diffuse non-filamentary gas forming a kind of disk shape and the other is filaments that are made up of finer threads forming an incomplete torus. Also, from the brightness fluctuations in the filaments, we can assume that at greater resolution more component threads will be visible. These filamentary structures neither have any particular shape nor resemble any AGN jets. However, the overall morphology has similarities with the toroids produced by shock compression in \citet{Enblin2002MNRAS.331.1011E}, which is discussed in more detail later. Next, we notice that along the north filament (Filament N) more similar strands are visible which also bend in a similar way forming a `Rib' like structure. We assume these are similar to Filament N but are seen in projection. We also see `Parallel Strands' perpendicular to `Filament E', seemingly stretching in parallel to each other with the very north one reaching almost to the `Ribs'. These observational features indicate them also being in projection. On the other hand, `Filament E' runs almost in parallel with the north-south stretch of the main filament on the west side. On the south, we see a near mirror image of the `Filament N', i.e., the `Filament S' which is connected with the `Torus' region with a peculiar `Bridge' like structure. We also notice the presence of diffuse non-filamentary components in most of the phoenix region. Furthermore, even within the small patch of the diffuse component in the south, finer `Threads' are present along with `Filament S'. 

%%%%%%%%%%%%%
\subsection{Multi-frequency morphological comparison}
In Fig. \ref{fig:radio_maps}, we present a side-by-side comparison of the radio phoenix at 148, 323, 700, and 1280 MHz. 
However, since the 148 MHz observation has a much lower angular resolution compared to the rest, we mainly focused on the other three images at a higher resolution.

Firstly, the overall morphology of the radio phoenix is similar at all four frequencies. Although the 148 MHz image lacks resolution, the filamentary regions are somewhat distinguishable from the diffuse non-filamentary part as they are brighter in comparison. In the $9\arcsec$ resolution 323 MHz image, the individual filaments are much more clearly visible than in the 148 MHz image. In fact, the primary filaments, a characteristic feature of the radio phoenix, are observed at the other three frequencies. We also notice some traces of the `plume' like features, seen in 323 MHz, in the 700 MHz image as well. These features are absent in the 1.28 GHz image, possibly because of radiative cooling at higher frequency. On the other hand, the non-filamentary regions i.e., the region `H' and the `Bay' area have low brightness at all frequencies. Furthermore, the apparent missing emission at the `Bay' at 323 and 1280 MHz images is presumably due to lack of sensitivity, as we see the presence of the same in the more sensitive 700 MHz image. 

Another noticeable feature is the `Arc' on the north-eastern side of the phoenix indicated previously in Fig. \ref{fig:325MHz_radio_map}. This low-brightness arc is clearly observed at 323 MHz (Fig. \ref{fig:Arc} \textit{left panel}) and already starts to fade out at 608 MHz (Fig. \ref{fig:Arc} \textit{middle panel}). What is interesting is that the `Arc' coincides with the shock associated with the infalling south-west subcluster. In Fig. \ref{fig:Arc} \textit{right panel}, we show a shock map derived from \textit{Chandra} X-ray observations (see \citealt{Rahaman2022MNRAS.515.2245R} for more details). Here, we see that the `Arc' roughly coincides with the incoming shock. However, since `Filament E' does not seem to be affected in a similar way, we speculate that the `Arc' is not associated with the phoenix and is observed in projection. 

As indicated in Fig. \ref{fig:700MHz_radio_map}, the southern part of the phoenix is connected with the central `Torus' by a single filament. To further explore this feature, we plotted brightness profiles corresponding to three different frequencies (i.e., 323. 700, and 1280 MHz in Fig. \ref{fig:radio_maps}) in two different directions. In Fig. \ref{fig:Bridge_profile} (\textit{left panel}) we see the expected brightness drop from the `Torus' to the southern part of the phoenix as the `Torus' is the brightest part. However, instead of a smooth and gradual decrease, we observe a dip in brightness around the `Bridge' region. Furthermore, in the \textit{middle panel}, we show a brightness profile across the filament and we see a `trough and ridge' like characteristic, as is clearly observed in Fig. \ref{fig:700MHz_radio_map} \& \ref{fig:radio_maps}. Since these characteristics are observed at three different frequencies (i.e., 323, 700, and 1280 MHz), it excludes the possibility of being an image artifact and indicates the southern part of the phoenix being loosely connected with the central torus. More specifically, although a filament from the torus connects the southern part, a slight brightness decrement happens for the diffuse component. Furthermore, we also observe a somewhat similar feature at the north end of the `Torus' which is clearly noticed in Fig. \ref{fig:700MHz_radio_map} as well as Fig. \ref{fig:radio_maps}. In Fig. \ref{fig:Bridge_profile} (\textit{right panel}), we plot a similar brightness profile as discussed before and we see a comparable characteristic for this region as well. Although we do not know the cause, it seems that the central `Torus' is a distinct feature in both filamentary and diffuse components of the phoenix.

%%%%%%%%%%%%
\subsection{Flux density estimation} \label{subsec:flux_density_est}
To estimate the flux density of the radio phoenix, it is necessary to subtract the contributions of the embedded point sources. 
However, in this case, all the compact radio sources (I, J, K, L, and M in Fig. \ref{fig:700MHz_radio_map}) are located around the periphery of the radio phoenix. 
Therefore, integrated flux densities of the phoenix are estimated by simply avoiding these compact sources. 
We derived flux densities of the radio phoenix at 148 (IM1), 323 (IM4), 700 (IM8), and 1280 (IM10) MHz to be $12.25\pm1.23$, $2.93\pm0.29$, $0.50\pm0.04$ and $0.077\pm0.004$ Jy, respectively. The flux densities are estimated within 3$\sigma_{\mathrm{rms}}$ of the images shown in Fig. \ref{fig:radio_maps}. 
Furthermore, these integrated flux density values only represent the total phoenix emission observed at respective frequencies, whereas the method to derive the integrated spectrum is discussed in Sect. \ref{subsec:Int_spectra}. Here, we would like to note that the total flux density contribution from the above compact sources is lower than the integrated flux density uncertainties of the radio phoenix at the respective frequencies. Therefore, even if we didn't avoid these point sources in our integrated flux density estimation, the above values would have remained largely unaffected.
The flux density uncertainties are calculated using 

\begin{equation}
    \sigma_{S} = \sqrt{(\sigma_{\mathrm{cal}}S)^2 + (\sigma_{\mathrm{rms}}\sqrt{N_{\mathrm{beam}}})^2}\,,
    \label{eq:flux_err}
\end{equation}

\noindent where $S$ is flux density, $\sigma_{\mathrm{cal}}$ and $\sigma_{\mathrm{rms}}$ are calibration uncertainty and map noise, respectively. 
The number of beams present within the 3$\sigma_{\mathrm{rms}}$ contour is denoted by $N_{\mathrm{beam}}$. 
We assumed the $\sigma_{\mathrm{cal}}$ corresponding to 148, 323, 700, and 1280 MHz to be $10\%$, $10\%$, $7\%$ \citep[][]{Chandra2004ApJ...612..974C} and $5\%$ \citep[e.g.,][]{Venturi2022A&A...660A..81V,Riseley2024A&A...686A..44R}, respectively, in the flux density error estimation.

\begin{figure*}
    \centering
    \begin{tabular}{cc}
    \includegraphics[width=\columnwidth]{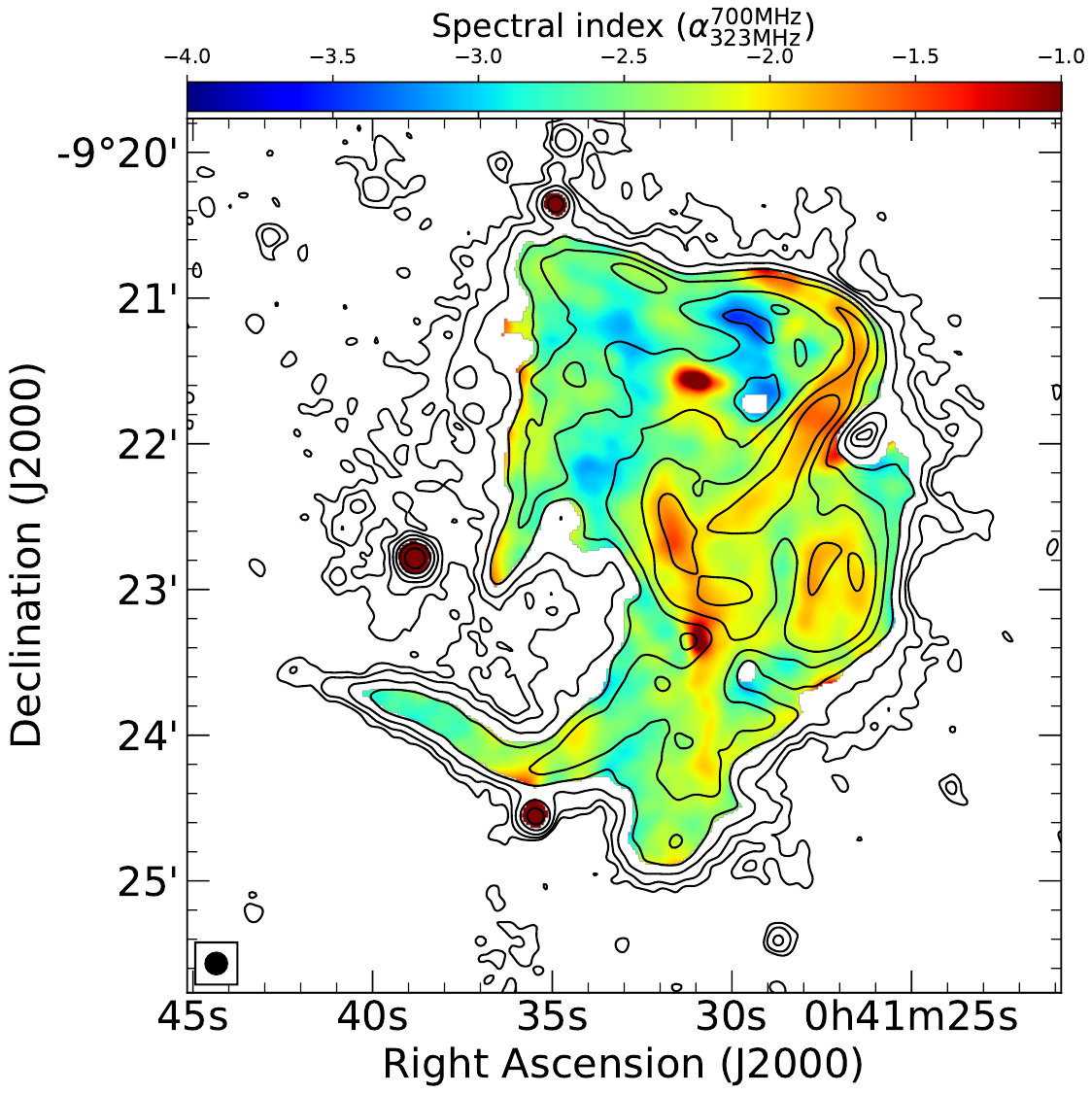} &
    \includegraphics[width=\columnwidth]{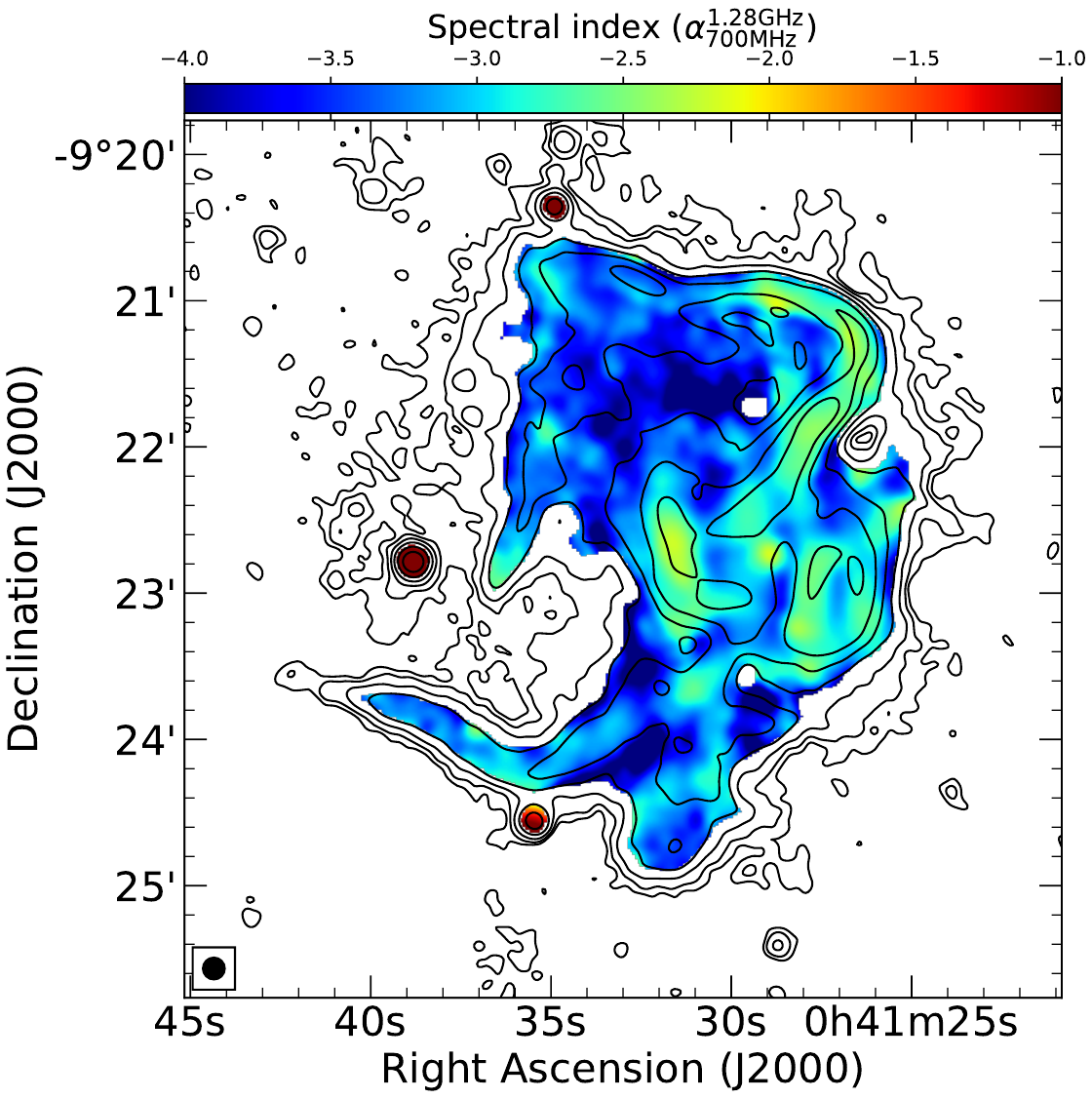} \\
    \includegraphics[width=\columnwidth]{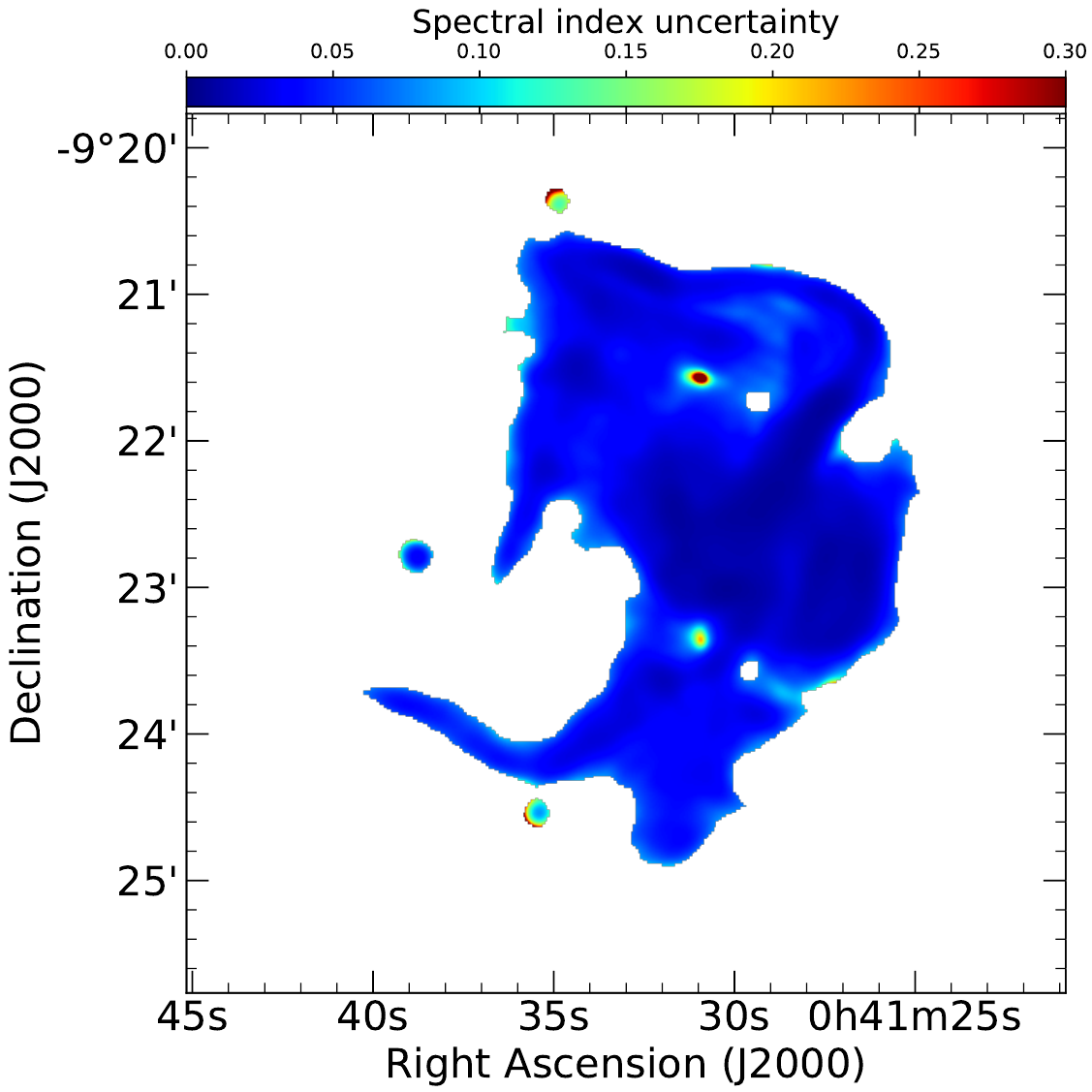} &
    \includegraphics[width=\columnwidth]{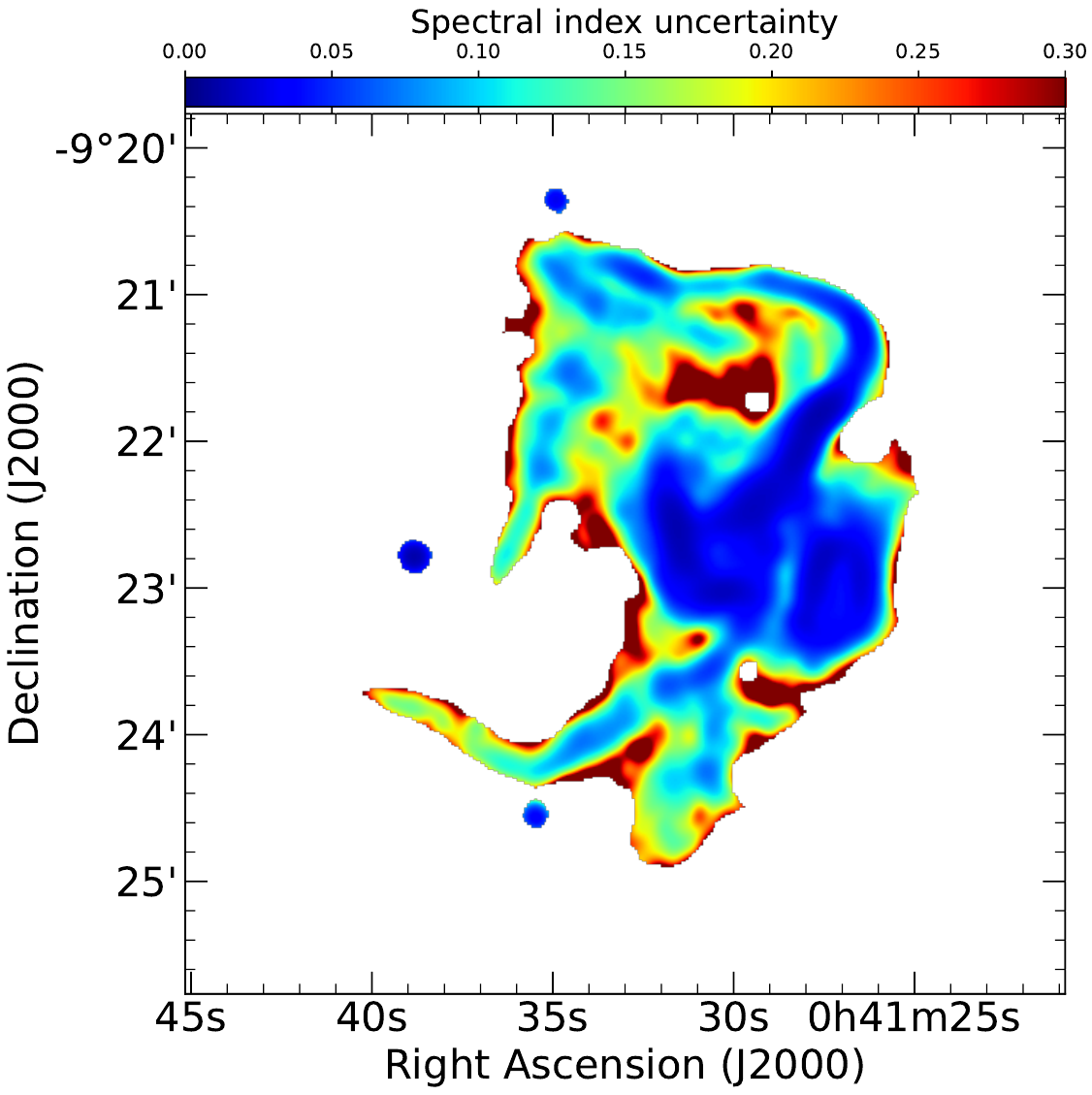} \\
    \end{tabular}
    \caption{\textit{Top row}: Spectral index maps between 323, 700, and 1280 MHz i.e., images IM4, IM8 and IM10. The angular resolution of the maps is $9\arcsec$ and is indicated at the bottom left corner. The contours overlaid on the maps correspond to the 700 MHz image (IM8) at levels $[1,2,4,8,...]\times 3\sigma_{\mathrm{rms}}$. 
    \textit{Bottom row}: Corresponding spectral index uncertainty map between 323 \& 700 MHz (\textit{left panel}) and 700 \& 1280 MHz \textit{(right panel)}.}
    \label{fig:spix_maps}
\end{figure*}

\begin{figure*}
    \centering
    \begin{tabular}{cc}
    \includegraphics[width=\columnwidth]{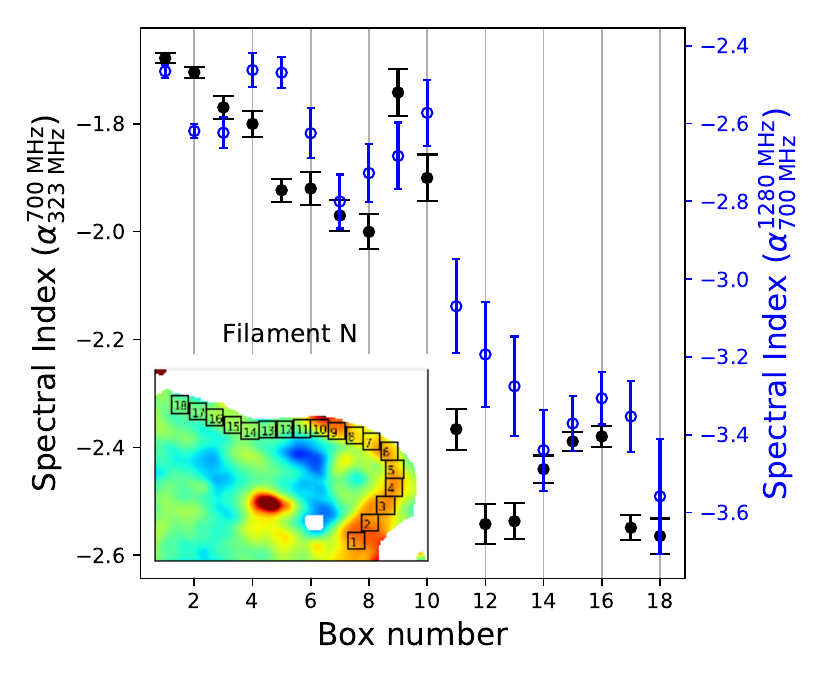}     &
    \includegraphics[width=\columnwidth]{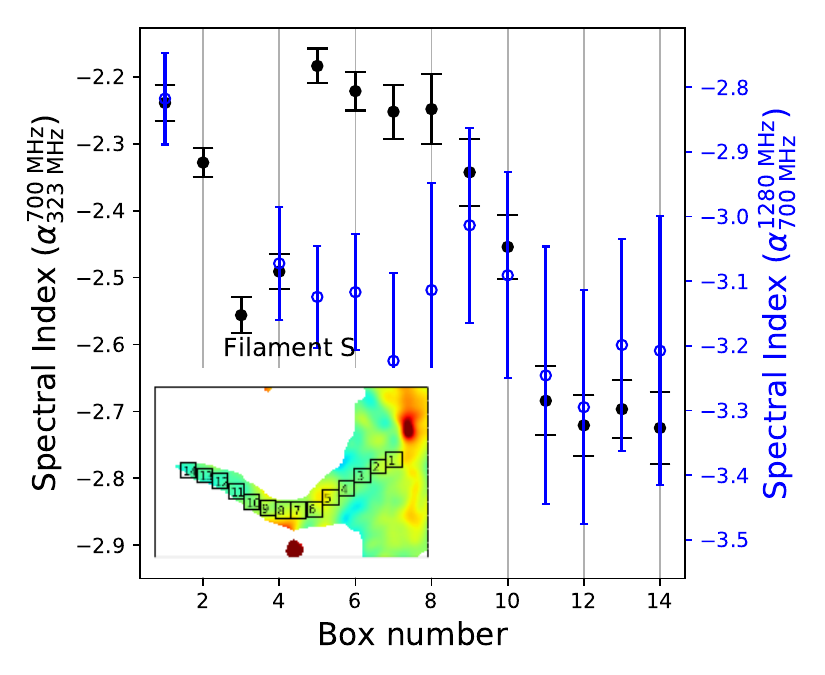} \\
    \end{tabular}
    \caption{Spectral index distribution along the `Filament N' (\textit{left}) and `Filament S' (\textit{right}) from 323 to 1280 MHz i.e., images IM4, IM8 and IM10. The data points represent the mean spectral index and uncertainty within the boxes. The size of the box regions is $9\arcsec$, the same as the beam size of the maps.}
    \label{fig:spix_fil_box}
\end{figure*}

\begin{figure*}
    \centering
    \begin{tabular}{cc}
    \includegraphics[width=\columnwidth]{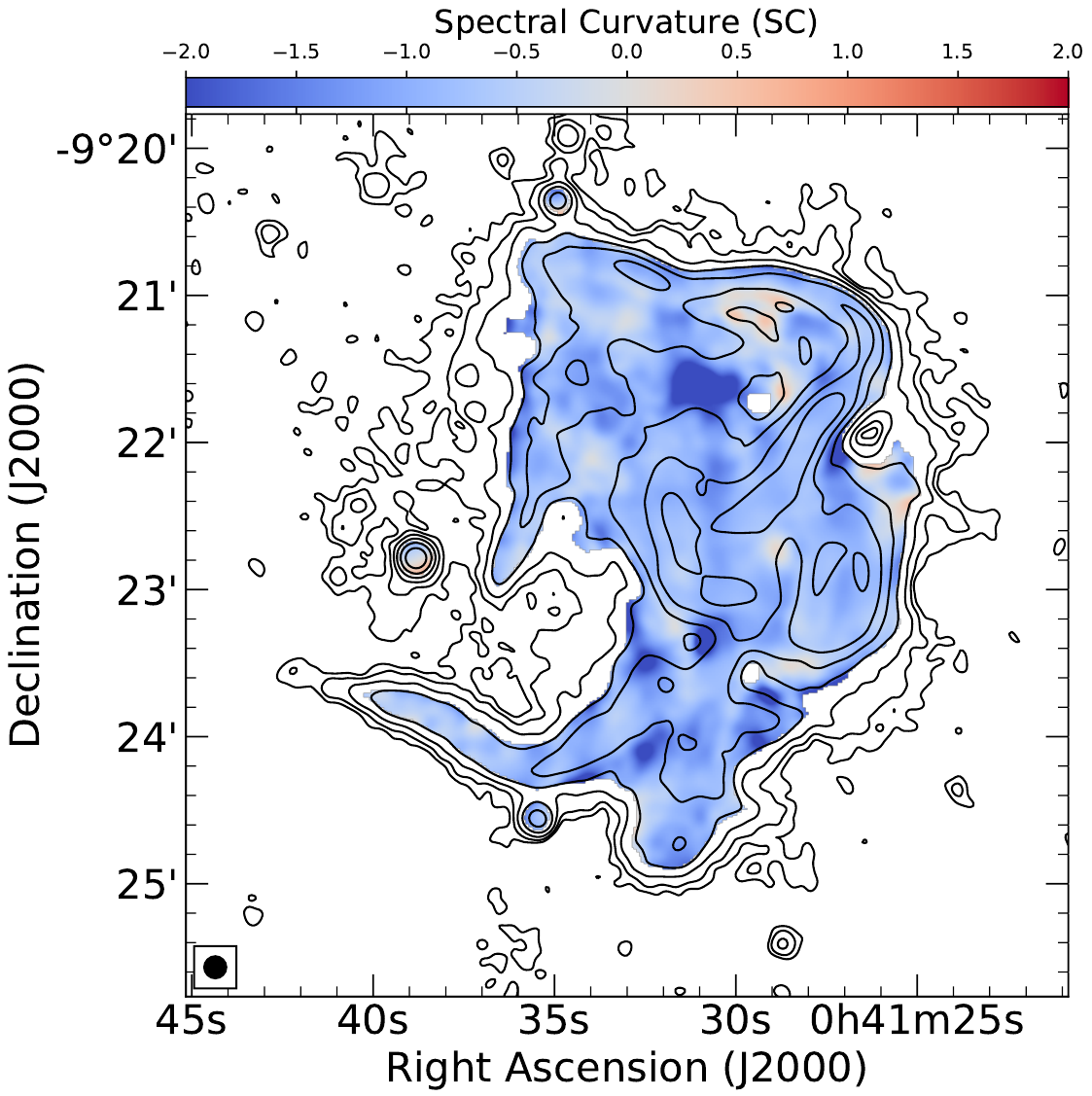}     &
    \includegraphics[width=\columnwidth]{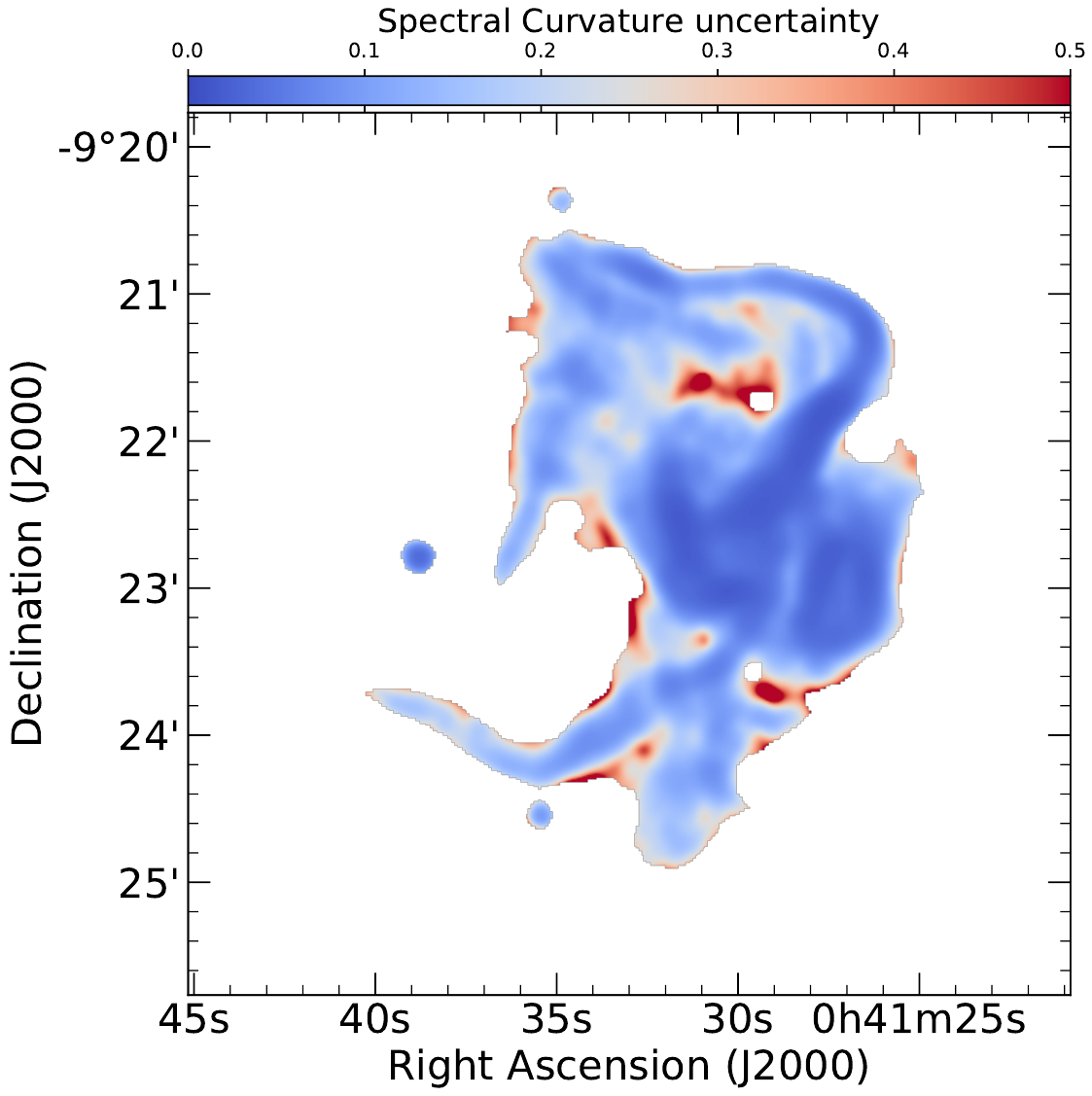}    \\
    \end{tabular}
    \caption{\textit{Left}: Spectral curvature map between 323, 700, and 1280 MHz i.e., images IM4, IM8 and IM10. The angular resolution of the map is $9\arcsec$ and is indicated at the bottom left corner. The contours overlaid on the maps correspond to the 700 MHz image (IM8) at levels $[1,2,4,8,...]\times 3\sigma_{\mathrm{rms}}$.  
    \textit{Right}: Corresponding spectral curvature uncertainty map.}
    \label{fig:curvature_map}
\end{figure*}

%%%%%%%%%%%%%%%%%%%%%%%%%%%%%%%%%
\section{Spectral study of the phoenix} \label{sec:spectral_study}
\subsection{Spectral index maps} \label{subsec:spix_map}

The Abell 85 cluster is observed at four different frequencies i.e., 148, 323, 700, and 1280 MHz.  
In Fig. \ref{fig:spix_maps}, we show two high-resolution ($9\arcsec$) spectral index maps derived from the 323 (IM4), 700 (IM8), and 1280 (IM10) MHz images presented in Fig. \ref{fig:radio_maps}. 
The spectral index uncertainty maps are produced using
\begin{equation}
    \Delta \alpha = \frac{1}{\ln{\big(\frac{\nu_1}{\nu_2}\big)}}\sqrt{\Bigg(\frac{\Delta S_1}{S_1}\Bigg)^2 + \Bigg(\frac{\Delta S_2}{S_2}\Bigg)^2}\,,
    \label{eq:spec_index_err}
\end{equation}
\noindent where $\nu$, $S$ and $\Delta S$ are image frequency, flux density value at each pixel, and map noise of the images, respectively.

The spectral index map between 323 \& 700 MHz in Fig. \ref{fig:spix_maps} (\textit{top left}) shows a significant fluctuation in the spectral index distribution. At first glance, we observe that most of the filamentary regions on the west side of the phoenix have flatter spectra (although in itself very steep $-1.5 \lesssim \alpha_{323}^{700} \lesssim -2$) compared to the other features ($-2.5 \lesssim \alpha_{323}^{700} \lesssim -3$); more specifically, the central `Torus' region and the western half of the `Filament N', which also correlates with them being the brightest part of the phoenix at all frequencies. We note that the spectral index uncertainty across the phoenix is small, mostly in the range of $\lesssim 0.05$ (Fig. \ref{fig:spix_maps} \textit{bottom left}). We also observe that along with the eastern half of `Filament N', `Filament S' and `Filament E' also have much steeper spectral indices. Furthermore, it is seen that the non-filamentary regions (e.g., region `H', `Bay') of the phoenix are spectrally the steepest part of the radio phoenix.

\begin{figure}
    \centering
    \includegraphics[width=\columnwidth]{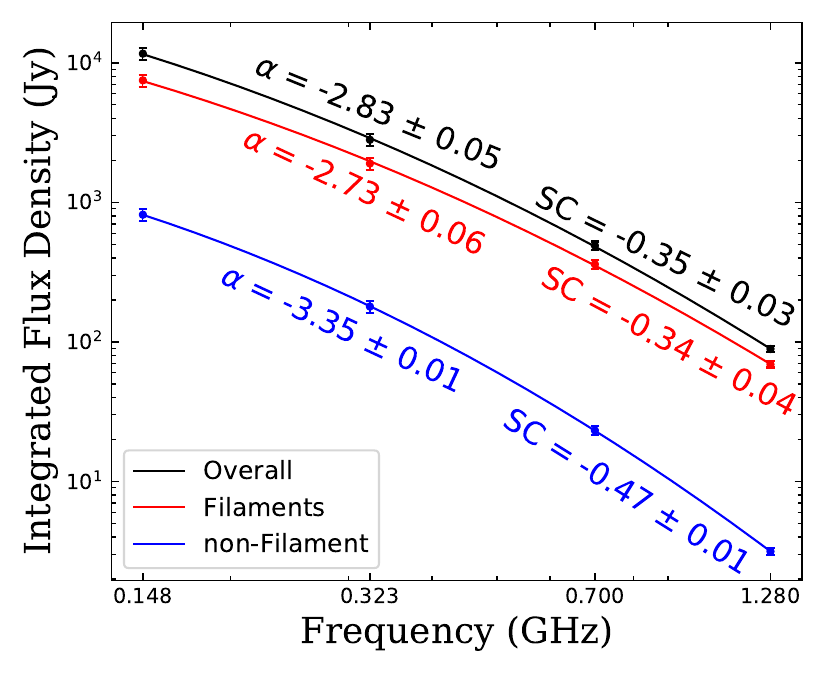}
    \caption{Integrated curved spectrum fit corresponding to different phoenix regions, where, $\alpha$ and SC are spectral index and curvature, respectively}. The images used for this purpose are IM2, IM5, IM9 and IM11. 
    \label{fig:spectra}
\end{figure}

Similar spectral fluctuations as observed on the previous map are present in the map between 700 \& 1280 MHz (Fig. \ref{fig:spix_maps} \textit{top right}) as well with the filamentary regions being more distinct. Here, the filamentary structures are more pronounced and closely resemble the brightness distribution as seen in Fig. \ref{fig:radio_maps}. A key difference is that the overall spectral index distribution has become much steeper than the previous map. This is observed as a spectral curvature in Fig. \ref{fig:curvature_map} \& \ref{fig:spectra} and is discussed later. The already steep filaments ($-1.5 \lesssim \alpha_{323}^{700} \lesssim-2$) show much steeper ($-2.3 \lesssim \alpha_{700}^{1280} \lesssim-2.8$) spectra. Furthermore, most of the non-filamentary regions have spectral indices steeper than $-3$, going as steep as $\sim -4$. On the uncertainty map (Fig. \ref{fig:spix_maps} \textit{bottom right}), we see that the filamentary regions have less spectral index uncertainty than the non-filamentary regions, in particular the edges and the `H' region. That is mostly due to very low brightness of those regions in the 1.28 GHz image, which reduces the signal to noise.

We observe in the spectral maps that the spectral index values are a bit flatter towards the western side of the filaments compared to the east, in particular the north and south filaments.
In Fig. \ref{fig:spix_fil_box} we plotted the spectral index distribution along `Filament N' and `Filament S' corresponding to the spectral index maps in Fig. \ref{fig:spix_maps}. We notice a clear spectral gradient along the filaments. The parts near the central torus are flatter compared to the far end at the east. The spectral gradient along `Filament N' is more evident in both maps owing to the filament being brighter at all frequencies. On the other hand, the less bright `Filament  S' shows a spectral gradient only between 323 \& 700 MHz and is obscured in the spectral map between 700 \& 1280 MHz because of higher spectral uncertainty. 

In Sect. \ref{sec:discus}, we further discuss these results in light of the existing formation mechanism.

%%%%%%%%%%%%%%%%%%%%%%%
\subsection{Spectral curvature map}

As we have multi-frequency radio data at more than two frequencies, we derived a spectral curvature map of the radio phoenix. Owing to the high resolution and good sensitivity of the radio images, we only used 323 to 1280 MHz data to produce a high spatially resolved spectral curvature map. The spectral curvature (SC) map is produced using radio images at three different frequencies or two spectral index maps as
\begin{equation}
    \mathrm{SC} = -\alpha^{\nu_1}_{\nu_2} + \alpha^{\nu_2}_{\nu_3}
    \label{eq:curvature}
\end{equation}
where $\alpha^{\nu_1}_{\nu_2}$ and $\alpha^{\nu_2}_{\nu_3}$ are the spectral indices derived previously using data from three frequencies \citep{Leahy1998ApJ...505..784L}.
Specifically, the low-frequency spectral index map ($\alpha^{\nu_1}_{\nu_2}$) used is between 323 \& 700 MHz and high-frequency map ($\alpha^{\nu_2}_{\nu_3}$) is between 700 \& 1280 MHz. The associated spectral curvature uncertainty is derived using equation \ref{eq:curvature_error}, and 
the resulting curvature map along with the uncertainty is presented in Fig. \ref{fig:curvature_map}. We notice that spectral curvature is negative in general throughout the map. Although some regions have curvature of around $-2$, mostly it is within $-0.5\lesssim \mathrm{SC} \lesssim-1.25$ with a few parts with positive curvature. Moreover, we do not observe any particular trend along the east-west direction as was previously observed in spectral index maps (Fig. \ref{fig:spix_fil_box}). 
A similar scenario was also reported by \citet{Elder2022A&A...666A...3E} in Abell 1033.
Although at first glance a gradual decrease in curvature is seen along `Filament S', the increasing uncertainty along with it makes this trend less robust. Furthermore, `Filament N', also being stretched in the same direction, does not show this kind of behavior. Additionally, apart from some local fluctuations, we do not observe any obvious curvature contrast between filaments and non-filamentary regions, as is observed in spectral index maps (Fig. \ref{fig:spix_maps}). 
Therefore, both the filamentary and non-filamentary components might have been ``energized" at the same event.
After that, synchrotron and inverse Compton loss have resulted in similar spectral curvature for both components. 
On the other hand, say, if one component was energized by the shock compression and the other component by some other method, which would have a different particle injection and subsequent radiative cooling time, resulting in a systematic structure in the curvature map. 
However, we observe a mostly homogeneous spectral curvature distribution with slight fluctuations here and there apart from the regions with larger uncertainties, irrespective of filamentary and non-filamentary regions (Fig. \ref{fig:sc_box}, \ref{fig:sc_plot}).
The adiabatic compression mechanism proposed by \citet{Enblin2001A&A...366...26E} predicts a steep and bent radio spectrum, which is in line with the observed curvature map.

%%%%%%%%%%%%%%%%%%%%%%%%%%%%%
\subsection{Integrated spectrum} \label{subsec:Int_spectra}
A steep and curved spectrum of the radio phoenix in Abell 85 is already shown in \citet{Enblin2001A&A...366...26E}, which also matches their proposed mechanism. 
In Fig. \ref{fig:spectra}, we have fitted a curved spectrum
to the integrated flux densities corresponding to different phoenix components.
We used a second-order polynomial fit of the form $y = ax^2 + bx +c$ where spectral index $\alpha = b$ and curvature $\mathrm{SC} = a$.
Here, the black line refers to the overall integrated spectrum of the radio phoenix, derived from the region where diffuse emission is present in all four frequencies (i.e., 148, 323, 700, and 1280 MHz). The spectrum has the characteristics of a typical radio phoenix, being very steep ($\alpha < -1.5$), and further steepening with frequency. We have also derived an integrated spectrum of the filamentary regions (red) consisting of the central bright torus along with north and south filaments and of the non-filamentary regions (blue) i.e., the `Bay' and `H' regions. 
Note that the flux density estimations were performed by avoiding the compact sources (wherever necessary) around the phoenix, as is previously discussed in Sect. \ref{subsec:flux_density_est}.
Similar to the spectral index maps, we see the filamentary region being less steep compared to the non-filamentary regions. 
Furthermore, the subtle curvature difference between filaments and non-filamentary region 
is largely biased because of inclusion of the `Bay’ region in the low resolution maps, which is clearly not part of the high resolution maps (Fig. \ref{fig:spix_maps}, \ref{fig:curvature_map}).

%%%%%%%%%%%%%%%%%%%%%%
\section{Deriving shock properties from the filaments} \label{sec:shock_prop}
To explain radio emission from the fossil plasma bubbles, \citet{Enblin2001A&A...366...26E} proposed that when a merging or accretion shock passes through the fossil plasma that is not older than 2 Gyr, it is compressed adiabatically. The energy gain during the compression along with the strengthening of the magnetic fields can revive the fossil plasma cocoon. A demonstration of the proposed model was performed by \citet{Enblin2002MNRAS.331.1011E} using hydrodynamical (MHD) simulations. The simulations show that the initial spherical cocoon is compressed and torn into filamentary structures, resulting in a toroidal shape in the presence of a weak magnetic field. 

Since an approximate measurement of the major $(R)$ and minor $(r)$ radii of the torus is possible from the sensitive high-resolution images presented in this work (i.e., the size of the radio phoenix and the filament widths as major and minor diameters of the torus), we can estimate the compression of the cocoon by the shock. Assuming the initial spherical radius of the radio plasma is similar to the major radius $(R)$ of the final torus, we can estimate the compression factor by
\begin{equation}
    C = \frac{V_\mathrm{sphere}}{V_\mathrm{torus}} = \frac{4\pi R^3/3}{2\pi^2 Rr^2} = \frac{2R^2}{3\pi r^2}.
\end{equation}
If the cocoon is in pressure equilibrium with its surroundings before and after the shock passage, then the pressure jump in the shock is $P_2/P_1 = C^{\gamma_\mathrm{rp}}$, where $\gamma_\mathrm{rp}$ is the adiabatic index of the radio plasma \citep{Enblin2002MNRAS.331.1011E}. From the pressure jump, we can calculate the Mach number using the shock jump condition \citep{Landau1959flme.book.....L}
\begin{equation}
    \frac{P_2}{P_1} = \frac{2\gamma M^2 - (\gamma -1)}{\gamma + 1},
\end{equation}
where $M$ is the Mach number. The shock properties of the radio phoenix are derived corresponding to different filament thicknesses (Fig. \ref{fig:filament_width}), observed in the high-resolution images shown above, and are presented in Table \ref{tab:Mach} \& \ref{tab:Mach2}. Here, the full phoenix size and the `Torus' only region will roughly represent the maximum and minimum size of the initial AGN plasma bubble. Even though the diffuse emission observed in Fig. \ref{fig:325MHz_radio_map} \& \ref{fig:700MHz_radio_map} does not have the ideal torus morphology, the ratio of its global or `Torus' diameter and filament thickness should correlate with the shock strengths derived here \citep{Enblin2002MNRAS.331.1011E}.

% Table
\begin{table}
\caption{Shock properties corresponding to the phoenix size}
\vspace{-0.2in}
\label{tab:Mach}
\begin{flushleft}
\begin{tabular}{|c|c|c|cc|cc|}
\hline
\multirow{2}{*}{Label} & \multirow{2}{*}{$d (\arcsec)$} & \multirow{2}{*}{$C$} & \multicolumn{2}{c|}{$\gamma_\mathrm{gas} = 5/3$}      & \multicolumn{2}{c|}{$\gamma_\mathrm{rp} = 4/3$}     \\ \cline{4-7} 
 &                      &                    & \multicolumn{1}{c|}{$P_2/P_1$}  & Mach   & \multicolumn{1}{c|}{$P_2/P_1$} & Mach   \\ \hline
A & 21.0                   & 33.75              & \multicolumn{1}{c|}{352.49} & 16.8 & \multicolumn{1}{c|}{109.07} & 9.78 \\ \hline
B & 24.6                   & 24.60              & \multicolumn{1}{c|}{208.01} & 12.91 & \multicolumn{1}{c|}{71.53}  & 7.92  \\ \hline
C & 23.3                   & 27.42               & \multicolumn{1}{c|}{249.28}  & 14.13  & \multicolumn{1}{c|}{82.67}  & 8.51  \\ \hline
D & 12.9                   & 89.44               & \multicolumn{1}{c|}{1788.87}  & 37.83  & \multicolumn{1}{c|}{400.00} & 18.71  \\ \hline
E & 11.6                   & 110.61               & \multicolumn{1}{c|}{2548.88}  & 45.16  & \multicolumn{1}{c|}{530.98} & 21.56  \\ \hline
F & 11.8                   & 106.90               & \multicolumn{1}{c|}{2407.71}  & 43.89  & \multicolumn{1}{c|}{507.32} & 21.07  \\ \hline
\end{tabular}
\end{flushleft}
\vspace{-0.3in}
\tablenotetext{}{\\Note: $d=2r$ is the filament width, $\gamma_\mathrm{gas}$ for ideal gas equation of state (monoatomic) and $\gamma_\mathrm{rp}$ for ultra-relativistic equation of state. Assumed size of the initial cocoon $D=2R \simeq 265$ arcsec i.e., the size of the radio phoenix.}
\end{table}

% Table
\begin{table}
\caption{Shock properties corresponding to the `Torus' size}
\vspace{-0.2in}
\label{tab:Mach2}
\begin{flushleft}
\begin{tabular}{|c|c|c|cc|cc|}
\hline
\multirow{2}{*}{Label} & \multirow{2}{*}{$d (\arcsec)$} & \multirow{2}{*}{$C$} & \multicolumn{2}{c|}{$\gamma_\mathrm{gas} = 5/3$}      & \multicolumn{2}{c|}{$\gamma_\mathrm{rp} = 4/3$}     \\ \cline{4-7} 
 &                      &                    & \multicolumn{1}{c|}{$P_2/P_1$}  & Mach   & \multicolumn{1}{c|}{$P_2/P_1$} & Mach   \\ \hline
A & 21.0                   & 5.63              & \multicolumn{1}{c|}{17.82} & 3.80 & \multicolumn{1}{c|}{10.02} & 2.98 \\ \hline
B & 24.6                   & 4.10              & \multicolumn{1}{c|}{10.52} & 2.93 & \multicolumn{1}{c|}{6.57}  & 2.42  \\ \hline
C & 23.3                   & 4.57               & \multicolumn{1}{c|}{12.60}  & 3.21  & \multicolumn{1}{c|}{7.59}  & 2.60  \\ \hline
D & 12.9                   & 14.92               & \multicolumn{1}{c|}{90.43}  & 8.52  & \multicolumn{1}{c|}{36.73} & 5.68  \\ \hline
E & 11.6                   & 18.45               & \multicolumn{1}{c|}{128.85}  & 10.16  & \multicolumn{1}{c|}{48.76} & 6.54  \\ \hline
F & 11.8                   & 17.83               & \multicolumn{1}{c|}{121.72}  & 9.88  & \multicolumn{1}{c|}{46.59} & 6.39  \\ \hline
\end{tabular}
\end{flushleft}
\vspace{-0.3in}
\tablenotetext{}{\\Note: $d=2r$ is the filament width, $\gamma_\mathrm{gas}$ for ideal gas equation of state (monoatomic) and $\gamma_\mathrm{rp}$ for ultra-relativistic equation of state. Assumed size of the initial cocoon $D=2R=\sqrt{90\times130} \simeq 108$ arcsec i.e., the size of the central `Torus'.}
\end{table}

%%%%%%%%%%%%%%%%%%%%%%%%%%%%%%%%%%%%%%%%%%%%%%%%%%%%%%%%%%%%%%%%%%%%%%%%%%%%%
\section{Comparison with the model} \label{sec:discus}
Classically, radio phoenices were classified together with radio relics and were treated as objects with similar origin scenarios. \citet{Kempner2004rcfg.proc..335K} then proposed to classify them separately based on the possible differences in the formation mechanism. 
The radio relics are proposed to be formed by diffusive shock re-acceleration mechanism \citep[e.g.,][]{Shimwell2015MNRAS.449.1486S,vanWeeren2017NatAs...1E...5V,Botteon2020A&A...634A..64B}, multiple shock acceleration \citep[e.g.,][]{Inchingolo2022MNRAS.509.1160I,Smolinski2023MNRAS.526.4234S}, whereas the origin mechanism of radio phoenices was proposed to be adiabatic compression of old AGN lobes in the shock \citep[][]{Enblin2001A&A...366...26E,Enblin2002MNRAS.331.1011E}. Here, we have performed a multi-frequency investigation of a radio phoenix with much higher sensitivity and resolution than previously reported.

This prototypical radio phoenix in the Abell 85 cluster shows an abundance of filaments (Fig. \ref{fig:325MHz_radio_map} \& \ref{fig:700MHz_radio_map}) with complex morphology as well as has a torus-like structure, as predicted in the simulations by \citet{Enblin2002MNRAS.331.1011E}. In fact, there are qualitative similarities between this radio phoenix and the simulation presented in \citet{Enblin2002MNRAS.331.1011E}. It was identified by \citet{Enblin2002MNRAS.331.1011E} as a double torus system with the central torus formed with bright filamentary structures in the central part. With the morphological comparison between Abell 85 phoenix and Fig. 6 of \citet{Enblin2002MNRAS.331.1011E}, they suggested that the phoenix environment has a weak magnetic field and the shock responsible was a weak one. Note that we employ the `weak' and `strong' labels here as per \citet{Enblin2002MNRAS.331.1011E}. They defined `weak' shock as corresponding to compression factor = 2 and `strong' to 3.3. Furthermore, dynamically unimportant magnetic fields are labeled as `weak' and dynamically important as `strong'.

Comparing the above phoenix images with Fig. 2 of \citet{Enblin2002MNRAS.331.1011E} indicates that the orientation of the filaments i.e., a northeast side open half-torus with the bow towards the southwest can form only when the shock passes through the cocoon from the northeast to southwest. This is similar to the merger-induced outgoing shocks, typically observed in the cluster periphery \citep{vanWeeren2012A&A...546A.124V,vanWeeren2017ApJ...835..197V,DiGennaro2018ApJ...865...24D}. In that case, the shock detected at the north-east end of the radio phoenix by \citet{Ichinohe2015MNRAS.448.2971I,Rahaman2022MNRAS.515.2245R} may not be the primary contributor to energizing the cocoon electrons. 
Furthermore, the spectral gradient from northeast to southwest observed in Fig. \ref{fig:spix_fil_box} is also an indicator of the shock passing in that direction.

Additionally, Fig. \ref{fig:spix_maps} highlights the brightness contrast between filamentary and non-filamentary regions. This may indicate the increase in gas density and magnetic field strength due to compression in the filamentary regions, as predicted by \citet{Enblin2002MNRAS.331.1011E}. In this scenario, when a shock passes through the fossil AGN bubble, a greater amount of the energy is injected into the filaments (formed in the process) compared to the rest of the medium, resulting in a less steep spectrum in the filaments compared to the non-filamentary regions.

The presence of most of the extended emission in the radio phoenix even at 1.28 GHz may indicate that the revival of the electron population is fairly recent. 
Furthermore, the spectral steepening of the phoenix (Fig. \ref{fig:spix_maps}, \ref{fig:curvature_map} \& \ref{fig:spectra}) highlights the low energy injection into the radio cocoon during the shock compression. Hence, even though the shock energization is fairly recent, the spectral steepening is already below GHz frequencies, as evident from the curved spectrum (Fig. \ref{fig:spectra}). This supports the shock strength being weak. 
Furthermore, the shock moving from northeast to southwest manifests as an ultra-steep spectral index on the northeastern side resulting from synchrotron loss after the shock passage, whereas the southwestern side is less steep, as the shock passage is more recent compared to the northeastern side. 
Hence, in accordance with the mechanism put forwarded by \citet{Enblin2001A&A...366...26E,Enblin2002MNRAS.331.1011E}, we propose that an outward-moving merger weak shock passed through an old AGN plasma (possibly related to one of the compact sources present in the region) with weak magnetic field strength, which resulted in the observed filamentary and torus morphology (Fig. 2 \& 6 of \citealt{Enblin2002MNRAS.331.1011E}) as well as the spectral gradient in the direction of the shock passage (Fig. \ref{fig:spix_fil_box}). 
Since the ICM density is much lower in the peripheral region of the cluster, it is difficult to detect the presence of an X-ray shock in the southwest of the radio phoenix. This requires much deeper X-ray observations than those that are presently available.

However, the above scenario does not completely explain the thin nature of the filaments in conjunction with the significant amount of extended diffuse gas present in this radio phoenix. Furthermore, the high compression factor and Mach numbers calculated from the filament widths suggest a strong shock (Table \ref{tab:Mach} \& \ref{tab:Mach2}). Besides, the presence of strong magnetic fields is needed to explain the diffuse non-filamentary component (Fig. 7, \citealt{Enblin2002MNRAS.331.1011E}), and a strong shock for the thin filaments and bright inner torus (Fig. 10, \citealt{Enblin2002MNRAS.331.1011E}). Furthermore, the presence of strong magnetic fields may amplify the radiative cooling, resulting in the observed spectral steepening and curvature of the phoenix (Fig. \ref{fig:curvature_map}, \ref{fig:spectra}). 
Moreover, the derived shock Mach numbers from a `Torus' size old AGN bubble give much more realistic values than the full phoenix size. Therefore, a scenario where a strongly magnetized initial plasma cocoon might have been around the size of the central `Torus', which was compressed with a strong outgoing shock, resulting in the observed radio phoenix with a complex structure.

%%%%%%%%%%%%%%%%%%%%%%%%%%%%%%%%%%%%%%%%%%%%%%%%%%%%%%%%%%%%%%%%%
\section{Summary and Conclusions} \label{sec:conclude}
In this work, we have performed a multi-frequency analysis of the radio phoenix in the Abell 85 cluster at 148, 323, 700, and 1280 MHz. With deep high-resolution images, a lot more detailed structures of the radio phoenix are revealed. 
We produced spectral index and curvature maps of the radio phoenix between 323, 700 \& 1280 MHz.
The torus filament orientation and the spectral gradient visible along the north and south filaments of the radio phoenix from northeast to southwest possibly correspond to the direction of the shock propagation. Furthermore, the spectral index distribution across the radio phoenix was found to be inhomogeneous, in particular, the filaments are less steep compared to the non-filamentary regions. 
Spectral curvature was found to be negative across most of the phoenix region, albeit with near uniform distribution regardless of the filaments. This may point towards the filamentary and non-filamentary components of the radio phoenix being associated with the same energization event.
We also derived shock properties from the filament widths of the phoenix following \citet{Enblin2002MNRAS.331.1011E}. 
All the features discussed above seem to be in support of the adiabatic compression by shock waves mechanism in the radio phoenix formation proposed by \citet{Enblin2001A&A...366...26E,Enblin2002MNRAS.331.1011E}. 
We discussed both the weak-magnetic field weak shock ($wSwB$) and strong-magnetic field strong shock ($sSsB$) scenarios \citep{Enblin2002MNRAS.331.1011E} for the radio phoenix and found the latter to be more plausible.
Accurate magnetic field mapping of the cluster ICM and the presence/absence of an X-ray shock near the southwest boundary of the phoenix will provide significant insight into the formation mechanism. Future sensitive high-resolution full-stokes radio observations at more frequencies will be able to present a much clearer picture of the phoenix-forming process. The current radio telescopes such as uGMRT, VLA, LOFAR, and MeerKAT already have the capability to explore these diffuse radio sources in detail and should be exploited in this endeavor. We also need more similar studies for other clusters known to host radio phoenix to get robust evidence in support of the said mechanism. In addition, deep X-ray observations are crucial in finding the shocks responsible for phoenix formation in the vicinity of these diffuse sources.

%% Also note that the akcnowlodgment environment does not support long amounts of text. If you have a lot of people and institutions to acknowledge, do not use this command. Instead, create a new \section{Acknowledgments}.
\begin{acknowledgments}
We would like to thank IIT Indore and Rhodes University for providing the necessary computing facilities for data analysis. We thank the staff of GMRT, who made these observations possible. GMRT is run by the National Centre for Radio Astrophysics of the Tata Institute of Fundamental Research. 
The MeerKAT telescope is operated by the South African Radio Astronomy Observatory, which is a facility of the National Research Foundation, an agency of the Department of Science and Innovation.
RR would like to thank Tiziana Venturi for valuable comments and suggestions.
RR and OS's research is supported by the South African Research Chairs Initiative of the Department of Science and Technology and the National Research Foundation (SARChI 81737).
MR would like to thank DST for INSPIRE fellowship program for financial support (IF160343). 
MR acknowledges support from the National Science and Technology Council of Taiwan (MOST 109- 2112-M-007-037-MY3; NSTC 112-2628-M-007-003-MY3).  
This research is supported by DST-SERB, through ECR/2017/001296 grant awarded to AD. 
\end{acknowledgments}


\vspace{5mm}
\facilities{GMRT, MeerKAT}


\software{\texttt{CASA} \citep{McMullin2007ASPC..376..127M}, 
\texttt{SPAM} \citep{Intema2009A&A...501.1185I,Intema2017A&A...598A..78I},
\texttt{WSCLEAN} \citep{offringa-wsclean-2014,Offringa2017MNRAS.471..301O}, \texttt{CARACal} \citep{Jozsa2020}, \texttt{TRICOLOUR} (https://github.com/ratt-ru/tricolour), \texttt{CUBICAL} (https://cubical.readthedocs.io/en/latest/),
\texttt{Astropy} \citep{AstropyCollaboration2013A&A...558A..33A,AstropyCollaboration2018AJ....156..123A}, \texttt{APLpy} \citep{aplpy}, \texttt{Matplotlib} \citep{matplotlib}   }



\appendix
\section{Spectral curvature uncertainty calculation}
Rewriting equation \ref{eq:curvature},
\begin{equation*}
\begin{split}
    SC &= \alpha^{\nu_2}_{\nu_3} - \alpha^{\nu_1}_{\nu_2} \\
       &= \frac{\ln{(\frac{S_2}{S_3})}}{\ln{(\frac{\nu_2}{\nu_3})}} - \frac{\ln{(\frac{S_1}{S_2})}}{\ln{(\frac{\nu_1}{\nu_2})}} \\
       &= \frac{\ln{(\frac{\nu_1}{\nu_2})} \ln{(\frac{S_2}{S_3})} - \ln{(\frac{\nu_2}{\nu_3})} \ln{(\frac{S_1}{S_2})}}{\ln{(\frac{\nu_1}{\nu_2})} \ln{(\frac{\nu_2}{\nu_3})}} \\
       &= \frac{\ln{(\frac{\nu_1}{\nu_2})} [\ln{(S_2)} - \ln{(S_3)}] - \ln{(\frac{\nu_2}{\nu_3})} [\ln{(S_1)} - \ln{(S_2)}]}{\ln{(\frac{\nu_1}{\nu_2})} \ln{(\frac{\nu_2}{\nu_3})}} \\
       &= \frac{- \ln{(\frac{\nu_2}{\nu_3})} \ln{(S_1)} + [\ln{(\frac{\nu_1}{\nu_2})} + \ln{(\frac{\nu_2}{\nu_3})}] \ln{(S_2)} - \ln{(\frac{\nu_1}{\nu_2})} \ln{(S_3)}}{\ln{(\frac{\nu_1}{\nu_2})} \ln{(\frac{\nu_2}{\nu_3})}} \\
       &= \frac{- \ln{(\frac{\nu_2}{\nu_3})} \ln{(S_1)} + \ln{(\frac{\nu_1}{\nu_3})} \ln{(S_2)} - \ln{(\frac{\nu_1}{\nu_2})} \ln{(S_3)}}{\ln{(\frac{\nu_1}{\nu_2})} \ln{(\frac{\nu_2}{\nu_3})}} \\
\end{split}
\end{equation*}
gives the form
\begin{equation*}
\begin{split}
    SC &= \frac{1}{d} [a \ln{(x)} + b \ln{(y)} + c \ln{(z)}]
\end{split}
\end{equation*}
where,
\begin{equation*}
    x = S_1,\ y = S_2,\ z = S_3
\end{equation*}
\begin{equation*}
    a = - \ln{\Big(\frac{\nu_2}{\nu_3}\Big)},\ b = \ln{\Big(\frac{\nu_1}{\nu_3}\Big)},\ c = - \ln{\Big(\frac{\nu_1}{\nu_2}\Big)},\ d = \ln{\Big(\frac{\nu_1}{\nu_2}\Big)} \ln{\Big(\frac{\nu_2}{\nu_3}\Big)}
\end{equation*}
Therefore, the curvature uncertainty
\begin{equation} \label{eq:curvature_error}
    \Delta SC = \frac{1}{d} \sqrt{\Bigg(\frac{a \Delta x}{x}\Bigg)^2 + \Bigg(\frac{b \Delta y}{y}\Bigg)^2 + \Bigg(\frac{c \Delta z}{z}\Bigg)^2}
\end{equation}

\begin{figure}
    \centering
    \includegraphics[width=\columnwidth]{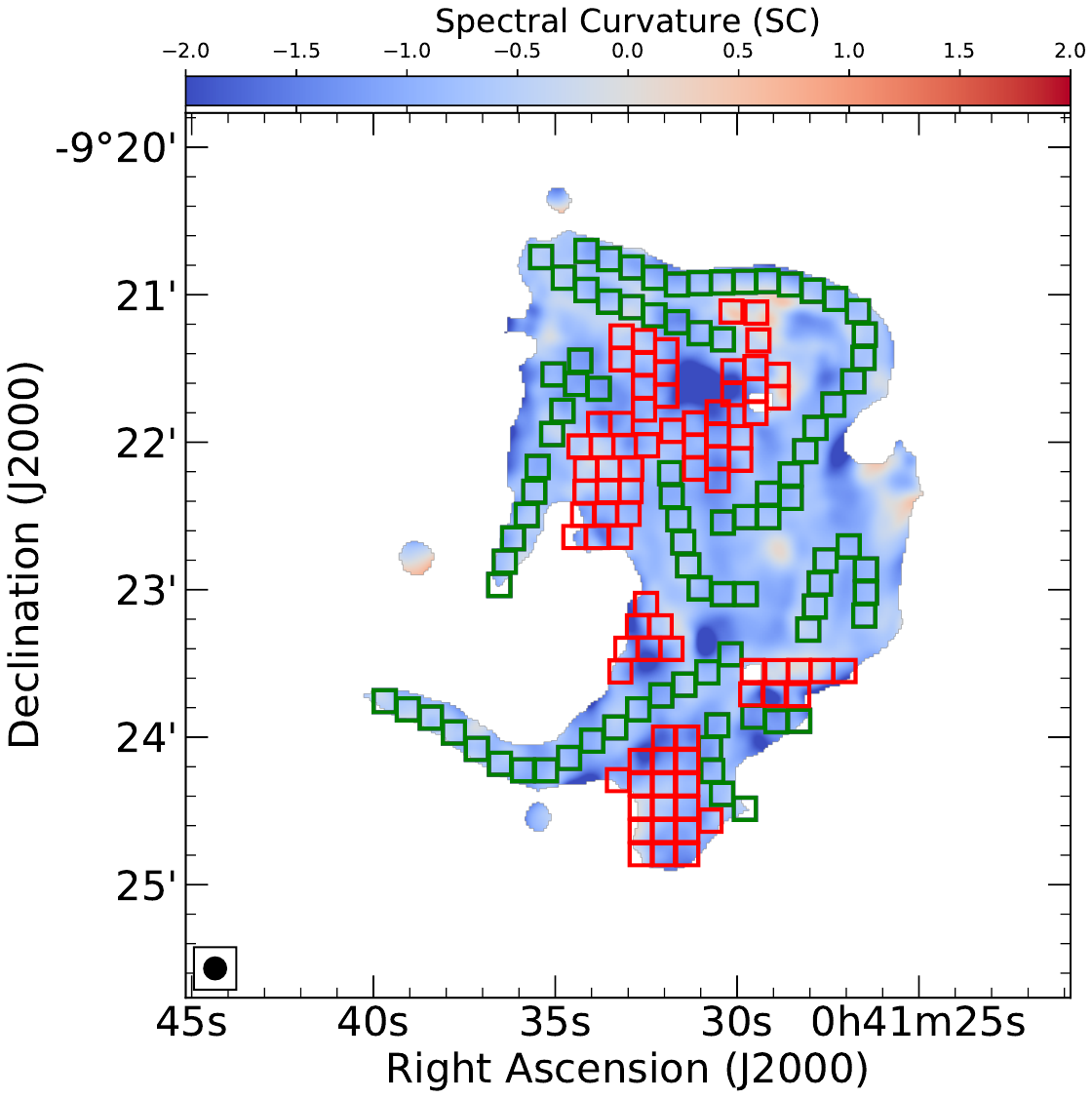}
    \caption{Box regions of $9\arcsec$ size for plotting spectral curvature distribution (Fig. \ref{fig:sc_plot}). Dark green boxes trace filaments, and red boxes cover non-filamentary regions.}
    \label{fig:sc_box}
\end{figure}

\begin{figure}
    \centering
    \includegraphics[width=\columnwidth]{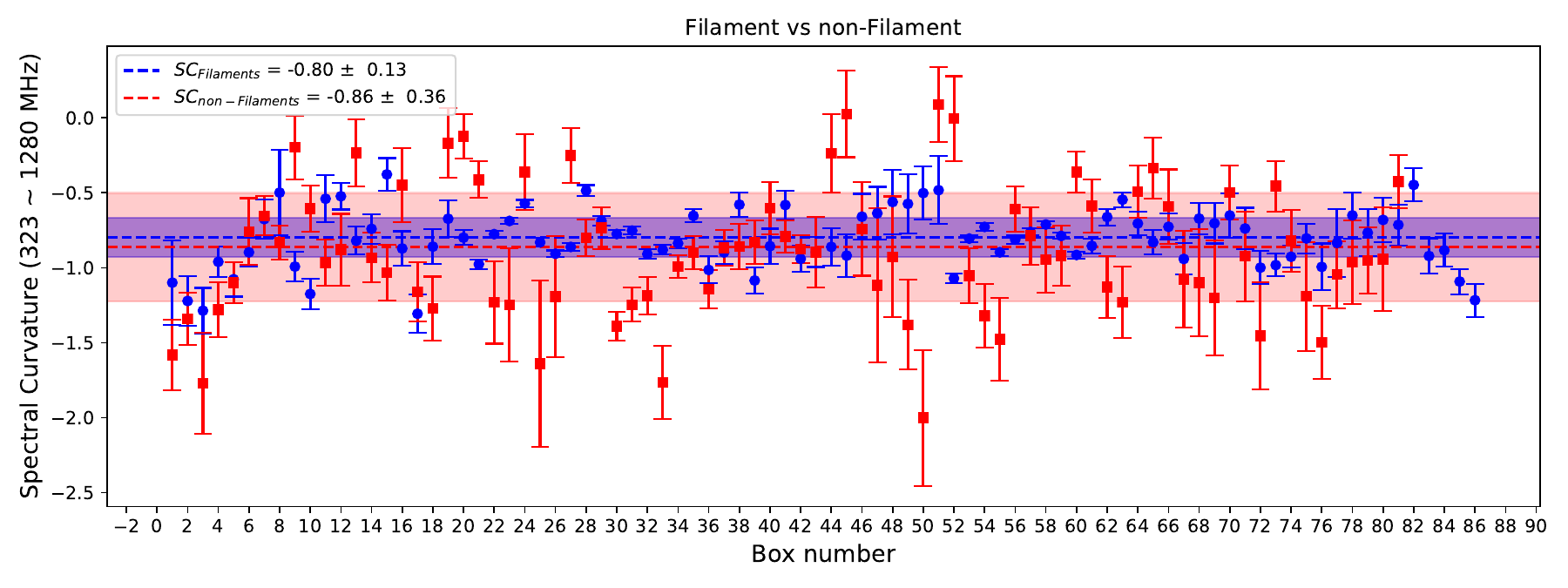}
    \caption{Spectral curvature distribution of filament and non-filamentary regions. The dashed lines and colored shaded regions show respective mean values and mean scatter, derived following \citet{Cassano2013ApJ...777..141C,vanWeeren2016ApJ...818..204V}.}
    \label{fig:sc_plot}
\end{figure}

\begin{figure}
    \centering
    \includegraphics[width=\columnwidth]{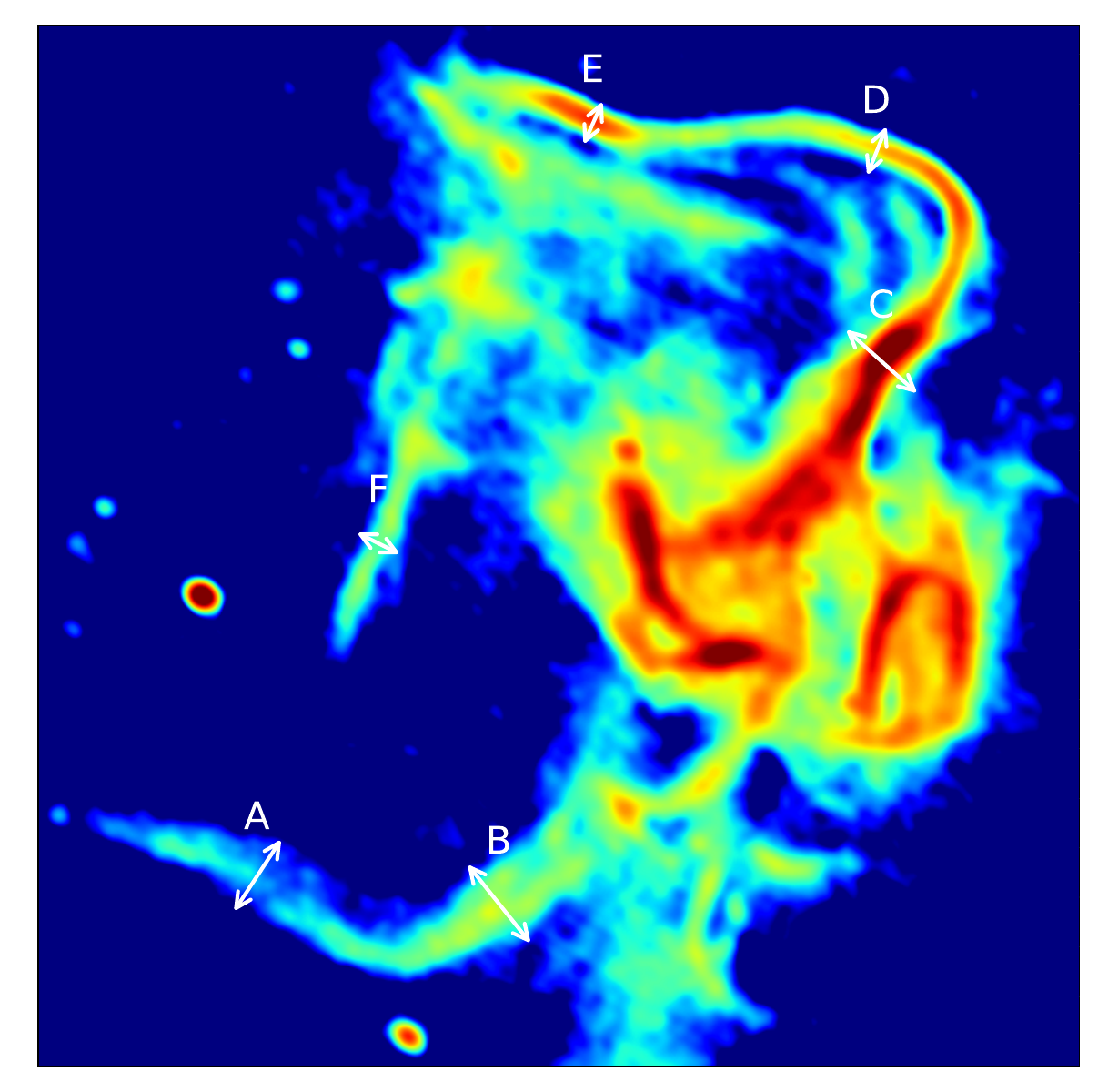}
    \caption{Labels where filament widths are measured. The image properties are given in Table \ref{tab:image_prop}, see label ``IM7''.}
    \label{fig:filament_width}
\end{figure}


\bibliography{myBib}{}
\bibliographystyle{aasjournal}



\end{document}